\documentclass[notitlepage,nofootinbib,superscriptaddress,showpacs,preprintnumbers,amsmath,amssymb,prd,onecolumn,12 point]{revtex4-1}
\usepackage{epsfig,subfigure}
\usepackage{graphicx}
\usepackage{bbm}
\usepackage{enumerate}
\usepackage[colorlinks=true,citecolor=blue]{hyperref}
\draft 
\begin{document}
\author{S. Dev}%
 \email{sdev@associates.iucaa.in}
  \affiliation{Department of Physics, School of Sciences, HNBG Central University, Srinagar, Uttarakhand 246174, INDIA.}%
  \author{Desh Raj}%
 \email{raj.physics88@gmail.com}
  \affiliation{Department of Physics, Himachal Pradesh University, Shimla 171005, INDIA.}%
\author{Radha Raman Gautam}%
 \email{gautamrrg@gmail.com}
  \affiliation{Department of Physics, Himachal Pradesh University, Shimla 171005, INDIA.}%
\title{Neutrino mass matrices with three or four vanishing cofactors and non diagonal charged lepton sector}

\begin{abstract}
We investigate the texture structures of lepton mass matrices with four (five) non-zero elements in the charged lepton mass matrix and three (four) vanishing cofactors in the neutrino mass matrix. Using weak basis transformations, all possible textures for three and four vanishing cofactors in $M_{\nu}$ are grouped into 7 classes and predictions for the unknown parameters such as the Dirac CP violating phase and the effective Majorana mass for the phenomenologically allowed textures have been obtained. We, also, illustrate how such texture structures can be realized using discrete Abelian flavor symmetries.
\end{abstract}
\maketitle
\section{INTRODUCTION}
Solar, atmospheric, and reactor neutrino experiments in addition to the more recent neutrino production from acceleration-based beams have provided some novel results over the last two decades or so and, invariably, strengthened the flavor Standard Model (SM). However, some critical ingredients like leptonic CP violation, neutrino mass hierarchy and neutrino masses are still missing. Furthermore, the nature of neutrinos (Majorana/Dirac) and absolute neutrino masses are still open issues. While the developments over the past two decades have brought out a coherent picture of neutrino mixing, the neutrino mass hierarchy, which is strongly correlated with the neutrino masses and the CP phase $\delta$ is still unknown. Specifically, the sign of $|\Delta m^{2}_{31}|=|m^{2}_{3}-m^{2}_{1}|$ is still unconstrained and is the focal issue for several ongoing and forthcoming experiments. In addition, recent neutrino oscillation data hint towards a non-maximal atmospheric mixing angle ($\theta_{23}$) which implies two possibilities: $\theta_{23} < \frac{\pi}{4}$ or $\theta_{23} > \frac{\pi}{4}$ \cite{th23} which when combined with the $\theta_{13}$-$\delta$ \cite{th13del} and the $\pm\Delta m^{2}_{31}$-$\delta$ degeneracy \cite{dms13del} leads to an overall eight fold degeneracy \cite{degeneracy}. In the SM, all fermion masses are Dirac masses which are generated via the Higgs mechanism. In order to have massive Dirac neutrinos, one has to necessarily enlarge the SM particle content by introducing right-handed neutrinos $\nu_{R}$. The $\nu_{L}$ and $\nu_{R}$ form a Dirac spinor $\Psi_{\nu}=\nu_{L}+\nu_{R}$ where $\nu_{R}$ are the additional spin states for the neutrinos. However, the gauge singlets $\nu_{R}$ can have a Majorana mass term $\nu_{R}^T C^{-1} M_{R} \nu_{R}$ where $M_{R}$, in general, is not diagonal in the flavor basis where $M_{D}$ is diagonal. Diagonalizing the full mass term leads to Majorana neutrinos and new mass eigenstates. Of course, one can attempt to forbid $M_{R}$ by postulating an additional symmetry such as lepton number conservation. \\
The mass matrix for Majorana neutrinos is, in general, complex symmetric containing nine physical parameters which include the three mass eigenvalues ($m_{1}, m_{2}, m_{3}$), the three mixing angles ($\theta_{13},\theta_{12}, \theta_{23}$) and the three CP-violating phases ($\alpha, \beta, \delta$). The two mass-squared differences ($\Delta m^{2}_{12}, \Delta m^{2}_{23}$) and the three neutrino mixing angles ($\theta_{12}, \theta_{23}, \theta_{13}$) have been measured in solar, atmospheric and reactor neutrino experiments. While the Dirac-type CP-violating phase $\delta$ will be probed in the forthcoming neutrino oscillation experiments the neutrinoless double beta decay searches will provide additional constraints on neutrino mass scale. The last unknown mixing angle $(\theta_{13})$ has been measured with a fairly good precision by a number of recent \cite{abe,adam,yabe,pfa,ahn} experiments. It is, however, clear that the neutrino mass matrix which encodes the neutrino properties has several unknown neutrino parameters which will remain undetermined even in the near future. Thus, the phenomenological approaches aimed at reducing the number of independent parameters are bound to play a crucial role in further development. There are several classes of predictive models in the literature such as texture zeros \cite{fram,dev,xing,branc,ahuja}, vanishing cofactors \cite{lashin,lavoura,gautam}, hybrid textures \cite{kan} and equality between elements \cite{lal} which explain the presently available neutrino oscillation data.\\ 
Neutrino mass matrices with texture zeros and vanishing cofactors are particularly interesting due to their connections to flavor symmetries. Neutrino mass models with texture zeros and vanishing cofactors have been widely studied in literature \cite{fram,dev,xing,branc,ahuja,lashin,lavoura,gautam,ludl,desh} for this reason. The lepton mass matrices with texture zeros and vanishing cofactors in both the charged lepton mass matrix $M_{l}$ and the neutrino mass matrix $M_{\nu}$ have been systematically studied in Refs. \cite{branc,ahuja,ludl}. Lepton mass matrices where the charged lepton mass matrix $M_{l}$ has four (five) non-zero elements while Majorana neutrino mass matrix $M_{\nu}$ have three (two) non-zero matrix elements have been studied recently in Ref. \cite{lavo} and an inverted neutrino mass ordering and a non-maximal Dirac-type CP-violating phase $\delta$ are predicted for all viable textures.  
The recent confirmation of a non-zero and not so small reactor mixing angle $\theta_{13}$ has emerged as an important discriminator of neutrino mass models and many models based on discrete symmetries have been discarded as these models require breaking of these symmetries to accommodate the current neutrino data.\\
In the present work, we consider new textures of lepton mass matrices and systematically, investigate their predictions for the unknown parameters. We show that these new textures can be realized on the basis of discrete Abelian $Z_{n}$ symmetries. Specifically, we investigate textures of lepton mass matrices with four (five) non-zero elements in $M_l$ and three (four) vanishing cofactors in $M_\nu$. The textures considered in the present work are as predictive as the textures with two texture zeros/two vanishing cofactors in the flavor basis. Moreover, vanishing cofactors in $M_{\nu}$ can be seen as zero entries in $M_{R}$ and $M_{D}$ within the framework of type-I seesaw mechanism \cite{seesaw}:
\begin{equation}
M_{\nu}=- M_{D} M_{R}^{-1} M_{D}^{T}.
\end{equation}
where $M_D$ is the Dirac neutrino mass matrix and $M_R$ is the right-handed Majorana neutrino mass matrix.\\
The texture structures related by weak basis transformations lead to the same predictions for neutrino parameters, hence, one cannot distinguish mass matrix structures related by weak basis transformations. The charged lepton mass matrix having 4 non-zero matrix elements with non-zero determinant (as none of the charged lepton masses is zero), can have the following form: 
\begin{equation}
M_{l}=\left(
\begin{array}{ccc}
\times & 0 & 0 \\
 0 & \times & \times \\
 0 & 0 & \times
\end{array}
\right).
\end{equation}
Other possible structures can be obtained by considering all possible reorderings of rows and columns of $M_{l}$. The Hermitian products $H_{l}=M_{l} M_{l}^{\dagger}$ corresponding to charged lepton mass matrices are given by 
\begin{eqnarray}
H_{l1}=\left(
\begin{array}{ccc}
m_{e}^{2} & 0 & 0 \\
 0 & \times & \times \\
 0 & \times & \times
\end{array}
\right),
H_{l2}=\left(
\begin{array}{ccc}
\times & 0 & \times \\
 0 & m_{\mu}^{2} & 0 \\
 \times & 0 & \times
\end{array}
\right),
H_{l3}=\left(
\begin{array}{ccc}
\times & \times & 0 \\
 \times & \times & 0 \\
 0 & 0 & m_{\tau}^{2}
\end{array}
\right)
\end{eqnarray}
where the diagonalization of $H_{l}$ gives the value of $V_{l}$ as $V_{l}^{\dagger} H_{l} V_{l}$=diag$(m_{e}^{2},m_{\mu}^{2},m_{\tau}^{2})$.\\
The neutrino mass matrices with three vanishing cofactors have following 20 distinct possible structures which have been classified into six classes:\\
Class-I
\begin{eqnarray}
M_{\nu 1}=\left(
\begin{array}{ccc}
\times & \Delta & \Delta \\
 \Delta & \times & \Delta \\
 \Delta & \Delta & \times
\end{array}
\right),
\end{eqnarray}\\
Class-II
\begin{eqnarray}
M_{\nu 2}=\left(
\begin{array}{ccc}
\Delta & \Delta &\times \\
\Delta & \times & \Delta \\
\times & \Delta & \times
\end{array}
\right),
M_{\nu 3}=\left(
\begin{array}{ccc}
\times & \Delta & \Delta \\
\Delta & \Delta & \times \\
\Delta & \times & \times
\end{array}
\right),
M_{\nu 4}=\left(
\begin{array}{ccc}
\Delta & \times & \Delta \\
\times & \times & \Delta \\
\Delta & \Delta & \times
\end{array}
\right),\nonumber \\
M_{\nu 5}=\left(
\begin{array}{ccc}
\times & \Delta & \Delta \\
\Delta & \times & \times \\
\Delta & \times & \Delta
\end{array}
\right),
M_{\nu 6}=\left(
\begin{array}{ccc}
\times & \times & \Delta \\
\times & \Delta & \Delta \\
\Delta &\Delta & \times
\end{array}
\right),
M_{\nu 7}=\left(
\begin{array}{ccc}
\times & \Delta &\times \\
\Delta & \times & \Delta \\
\times & \Delta & \Delta
\end{array}
\right),
\end{eqnarray}\\
Class-III
\begin{eqnarray}
M_{\nu 8}=\left(
\begin{array}{ccc}
\times & \times & \Delta \\
\times & \Delta & \times \\
\Delta & \times & \Delta
\end{array}
\right),
M_{\nu 9}=\left(
\begin{array}{ccc}
\times & \Delta &\times \\
\Delta & \Delta & \times \\
\times & \times & \Delta
\end{array}
\right),
M_{\nu 10}=\left(
\begin{array}{ccc}
\Delta & \times &\times \\
\times & \times & \Delta \\
\times & \Delta & \Delta
\end{array}
\right),\nonumber \\
M_{\nu 11}=\left(
\begin{array}{ccc}
\Delta & \Delta &\times \\
\Delta & \times & \times \\
\times & \times & \Delta
\end{array}
\right),
M_{\nu 12}=\left(
\begin{array}{ccc}
\Delta & \times & \times \\
\times & \Delta & \Delta \\
\times & \Delta & \times
\end{array}
\right),
M_{\nu 13}=\left(
\begin{array}{ccc}
\Delta & \times & \Delta \\
\times & \Delta & \times \\
\Delta & \times & \times
\end{array}
\right),
\end{eqnarray}\\
Class-IV
\begin{eqnarray}
M_{\nu 14}=\left(
\begin{array}{ccc}
\Delta & \Delta & \Delta \\
\Delta & \times & \times \\
\Delta & \times& \times
\end{array}
\right),
M_{\nu 15}=\left(
\begin{array}{ccc}
\times & \times & \Delta \\
 \times & \times & \Delta \\
 \Delta & \Delta & \Delta
\end{array}
\right),
M_{\nu 16}=\left(
\begin{array}{ccc}
\times & \Delta &\times \\
\Delta & \Delta & \Delta \\
\times & \Delta & \times
\end{array}
\right),
\end{eqnarray}\\
Class-V
\begin{eqnarray}
M_{\nu 17}=\left(
\begin{array}{ccc}
\Delta & \times &\times \\
\times & \Delta & \times \\
\times & \times & \Delta
\end{array}
\right),
\end{eqnarray}\\
Class-VI
\begin{eqnarray}
M_{\nu 18}=\left(
\begin{array}{ccc}
\times & \times &\times \\
\times & \Delta & \Delta \\
\times & \Delta & \Delta
\end{array}
\right),
M_{\nu 19}=\left(
\begin{array}{ccc}
\Delta & \times &\Delta \\
\times & \times & \times \\
\Delta & \times & \Delta
\end{array}
\right),
M_{\nu 20}=\left(
\begin{array}{ccc}
\Delta & \Delta &\times \\
\Delta & \Delta & \times \\
\times & \times & \times
\end{array}
\right)
\end{eqnarray}
where $\Delta$ at $(i,j)$ position represents vanishing cofactor corresponding to element $(i,j)$ while $\times$ denotes a non-zero arbitrary entry. Neutrino mass matrices in each class are related by $S_{3}$ permutation symmetry as $M_{\nu}\rightarrow S^{T} M_{\nu} S$, where $S$ denotes the permutation matrices corresponding to $S_{3}$ group:
\begin{eqnarray}
S_{1}&=&\left(
\begin{array}{ccc}
1 & 0 & 0 \\
0 & 1 & 0 \\
0 & 0 & 1
\end{array}
\right),
S_{123}=\left(
\begin{array}{ccc}
0 & 0 & 1 \\
1 & 0 & 0 \\
0 & 1 & 0
\end{array}
\right),
S_{132}=\left(
\begin{array}{ccc}
0 & 1 & 0 \\
0 & 0 & 1 \\
1 & 0 & 0
\end{array}
\right),\nonumber \\
S_{12}&=&\left(
\begin{array}{ccc}
0 & 1 & 0 \\
1 & 0 & 0 \\
0 & 0 & 1
\end{array}
\right),
S_{13}=\left(
\begin{array}{ccc}
0 & 0 & 1 \\
0 & 1 & 0 \\
1 & 0 & 0
\end{array}
\right),
S_{23}=\left(
\begin{array}{ccc}
1 & 0 & 0 \\
0 & 0 & 1 \\
0 & 1 & 0
\end{array}
\right).
\end{eqnarray}
\section*{Analysis}

The mass term for charged leptons and Majorana neutrinos can be written as
\begin{equation}
-\mathcal{L}_{mass}=\overline{l}_{L} M_{l} l_{R}- \frac{1}{2} \nu^{T}_{L} C^{-1}M_{\nu} \nu_{L}+H.c.
\end{equation}
where $C$ is the charge conjugation matrix. The charged lepton and the Majorana neutrino mass matrix can be diagonalized as  
\begin{equation}
V_{l}^{\dagger} M_{l}M_{l}^{\dagger} V_{l} = (M_{l}^{D})^2,~~~~~~~~~~V_{\nu}^{T} M_{\nu} V_{\nu} = M_{\nu}^{D}
\end{equation}
where $M^{D}_{l}=$ diag$(m_{e},m_{\mu},m_{\tau})$, and $M^{D}_{\nu}=$ diag$(m_{1},m_{2},m_{3})$. $V_{l}$ and $V_{\nu}$ are unitary matrices connecting mass eigenstates to the flavor eigenstates. The lepton mixing matrix also known as the Pontecorvo-Maki-Nakagawa-Sakata (PMNS) matrix \cite{pmns} is given by
\begin{equation}
U=V_{l}^{\dagger} V_{\nu}
\end{equation}
which can be parametrized in terms of three mixing angles and three CP-violating phases in the standard parametrization as \cite{pdgpara}
\begin{equation}
U=\left(
\begin{array}{ccc}
 c_{12} c_{13} & c_{13} s_{12} & e^{-i \delta } s_{13} \\
 -c_{23} s_{12}-e^{i \delta } c_{12} s_{13} s_{23} & c_{12} c_{23}-e^{i \delta } s_{12} s_{13} s_{23}
   & c_{13} s_{23} \\
 s_{12} s_{23}-e^{i \delta } c_{12} c_{23} s_{13} & -e^{i \delta }
  c_{23} s_{12} s_{13}-c_{12} s_{23} & c_{13} c_{23}
\end{array}
\right)~
\left(
\begin{array}{ccc}
1 & 0 & 0 \\
0 & e^{i \alpha}&0 \\
0 & 0 & e^{i (\beta+\delta)}
\end{array}
\right)
\end{equation}
where $c_{ij}=\cos \theta_{ij}$ and $s_{ij}=\sin \theta_{ij}$. $\alpha$ and $\beta$ are the two Majorana CP-violating phases and $\delta$ is the Dirac-type CP-violating phase.\\
The CP violation in neutrino oscillation experiments can be reflected in terms of Jarlskog rephasing invariant quantity $J_{CP}$ \cite{jarl} with
\begin{equation}
J_{CP}=\textrm{Im}\lbrace U_{11} U_{22} U^{*}_{12} U^{*}_{21}\rbrace=\sin\theta_{12} \sin\theta_{23} \sin\theta_{13} \cos\theta_{12} \cos\theta_{23} \cos^{2}\theta_{13} \sin\delta .
\end{equation}
The effective Majorana neutrino mass $|m_{ee}|$, which determines the rate of neutrinoless double beta decay is given by
\begin{equation}
|m_{ee}|=|m_1 U^{2}_{e1}+m_2 U^{2}_{e2}+m_3 U^{2}_{e3}|.
\end{equation}
There are a large number of experiments such as CUORICINO \cite{cuor}, CUORE \cite{arna}, MAJORANA \cite{major}, SuperNEMO \cite{nemo}, EXO \cite{danil} which aim to achieve a sensitivity up to $0.01 eV$ for $|m_{ee}|$.\\
Recent cosmological observations provide more stringent constraints on absolute neutrino mass scale. Planck satellite data \cite{planck} combined with WMAP, cosmic microwave background and baryon acoustic oscillation experiments limit the sum of neutrino masses $\sum\limits_{i=1}^3 m_{i}\leq 0.23$ eV at 95$\%$ confidence level (CL).

\section{Numerical Analysis}
In this section we present detailed numerical analysis along with the main predictions for viable textures. The charged lepton mixing matrices corresponding to structures $H_{l1}, H_{l2}, H_{l3}$ are given by
\begin{small}
\begin{eqnarray}
V_{l1}&=&\left(
\begin{array}{ccc}
1 & 0 & 0 \\
0 & \cos \theta & e^{i \phi_l} \sin \theta \\
0 & - e^{-i \phi_l} \sin \theta & \cos \theta
\end{array}
\right),\nonumber \\
V_{l2}&=&\left(
\begin{array}{ccc}
\cos \theta & 0 & e^{i \phi_l} \sin \theta \\
0 & 1 & 0  \\
- e^{-i \phi_l} \sin \theta & 0 & \cos \theta
\end{array}
\right),\nonumber \\
V_{l3}&=&\left(
\begin{array}{ccc}
\cos \theta & e^{i \phi_l} \sin \theta & 0 \\
- e^{-i \phi_l} \sin \theta & \cos \theta & 0 \\
0 & 0 & 1
\end{array}
\right)
\end{eqnarray}
\end{small}
respectively, with $\cos\theta=\sqrt{\frac{m_{y}-m}{m_{y}-m_{x}}}$. The parameters $m_{x}, m_{y}$ are defined as $m_{x}=m_{\mu}^{2}$, $m_{y}=m_{\tau}^{2}$ for the structure $H_{l1}$, $m_{x}=m_{e}^{2}$, $m_{y}=m_{\tau}^{2}$ for structure $H_{l2}$ and $m_{x}=m_{e}^{2}$, $m_{y}=m_{\mu}^{2}$ for structure $H_{l3}$. Also, the parameter $m$ is constrained as $m_{x}<m<m_{y}$. The charged lepton mass eigenvalues are $m_{e}=0.510998928$ MeV, $m_{\mu}=105.6583715$ MeV and $m_{\tau}=1776.86$ MeV \cite{pdg}.\\
Neutrino mass matrices of Class-I and Class-II lead to one or more zeros in the lepton mixing matrix $U$ and, hence, both classes are phenomenologically excluded. Neutrino mass matrices of Class-VI lead to two degenerate neutrino mass eigenvalues, which is inconsistent with the current experimental data and hence this class is, also, phenomenologically ruled out. Therefore, we focus on the other three non trivial classes i.e., Class-III, IV and V.\\
The 10 possible structures for $M_{\nu}$ from Classes-III, IV, V along with 3 structures for $H_{l}$, lead to a total of $10 \times 3=30$ possible combinations of charged lepton and neutrino mass matrices. But all possible combinations for $H_{l}$ and $M_{\nu}$ are not independent of each other as the transformations $M_{\nu}\rightarrow S M_{\nu} S^{T}$ and $H_{l}\rightarrow S H_{l} S^{\dagger}$ relate some of the textures with each other. Table \ref{tab1} contains all possible independent texture structure of $M_{\nu}$ and $H_{l}$ and their viabilities for Normal mass Ordering (NO) and Inverted mass Ordering (IO). 
\begin{table}[h]
\begin{center}
\begin{tabular}{|c|c|c|c|}
 \hline
Class & Texture & $H_{l}, M_{\nu}$ & Viability \\
&   &  & NO IO \\
 \hline 
 & III-(A)& $H_{l1}$, $M_{\nu 8} \sim H_{l1}$, $M_{\nu 9}$ & $\times~~~\times$  \\

 & III-(B)&  $H_{l2}$, $M_{\nu 10} \sim H_{l2}$, $M_{\nu 11}$ & $\times~~~\times$ \\
 
 & III-(C)& $H_{l3}$, $M_{\nu 13} \sim H_{l3}$, $M_{\nu 12}$ & $\times~~~\times$ \\
 
 & III-(D)& $H_{l1}$, $M_{\nu 10} \sim H_{l1}$, $M_{\nu 12}$ & $\times~~~\times$ \\
 
III & III-(E)& $H_{l2}$, $M_{\nu 8} \sim H_{l2}$, $M_{\nu 13}$ & $\times~~~\times$ \\

 & III-(F)& $H_{l3}$, $M_{\nu 11} \sim H_{l3}$, $M_{\nu 9}$ & $\times~~~\times$ \\

 & III-(G)& $H_{l1}$, $M_{\nu 11} \sim H_{l1}$, $M_{\nu 13}$ & $\times~~~\times$ \\
 
 & III-(H)& $H_{l3}$, $M_{\nu 8} \sim H_{l3}$, $M_{\nu 10}$ & $\surd~~~\times$ \\
 
 & III-(I)& $H_{l2}$, $M_{\nu 9} \sim H_{l2}$, $M_{\nu 12}$ & $\surd~~~\times$\\
 
 \hline
 & IV-(A)& $H_{l1}$, $M_{\nu 14}$ & $\times~~~\times$ \\
 
 & IV-(B)& $H_{l2}$, $M_{\nu 16}$ & $\times~~~\times$ \\
 
 & IV-(C)& $H_{l3}$, $M_{\nu 15}$ & $\times~~~\times$  \\

 IV & IV-(D)& $H_{l3}$, $M_{\nu 14} \sim H_{l3}$, $M_{\nu 16}$ & $\surd~~~\surd$ \\
 
 & IV-(E)& $H_{l2}$, $M_{\nu 14} \sim H_{l2}$, $M_{\nu 15}$ & $\surd~~~\surd$ \\
 
 & IV-(F)& $H_{l1}$, $M_{\nu 16} \sim H_{l1}$, $M_{\nu 15}$ & $\surd~~~\surd$ \\
 \hline 
 & V-(A)& $H_{l1}$, $M_{\nu 17}$ & $\times~~~\times$ \\
 
V & V-(B)& $H_{l2}$, $M_{\nu 17}$ & $\times~~~\times$ \\

 & V-(C)& $H_{l3}$, $M_{\nu 17}$ & $\times~~~\times$ \\
 \hline 
 \end{tabular}
\caption{Possible independent structures of $H_{l}, M_{\nu}$ and their viability.}
\label{tab1}
\end{center}
\end{table}

\subsection*{Class-III}

\begin{figure}[h]
\begin{center}
\epsfig{file=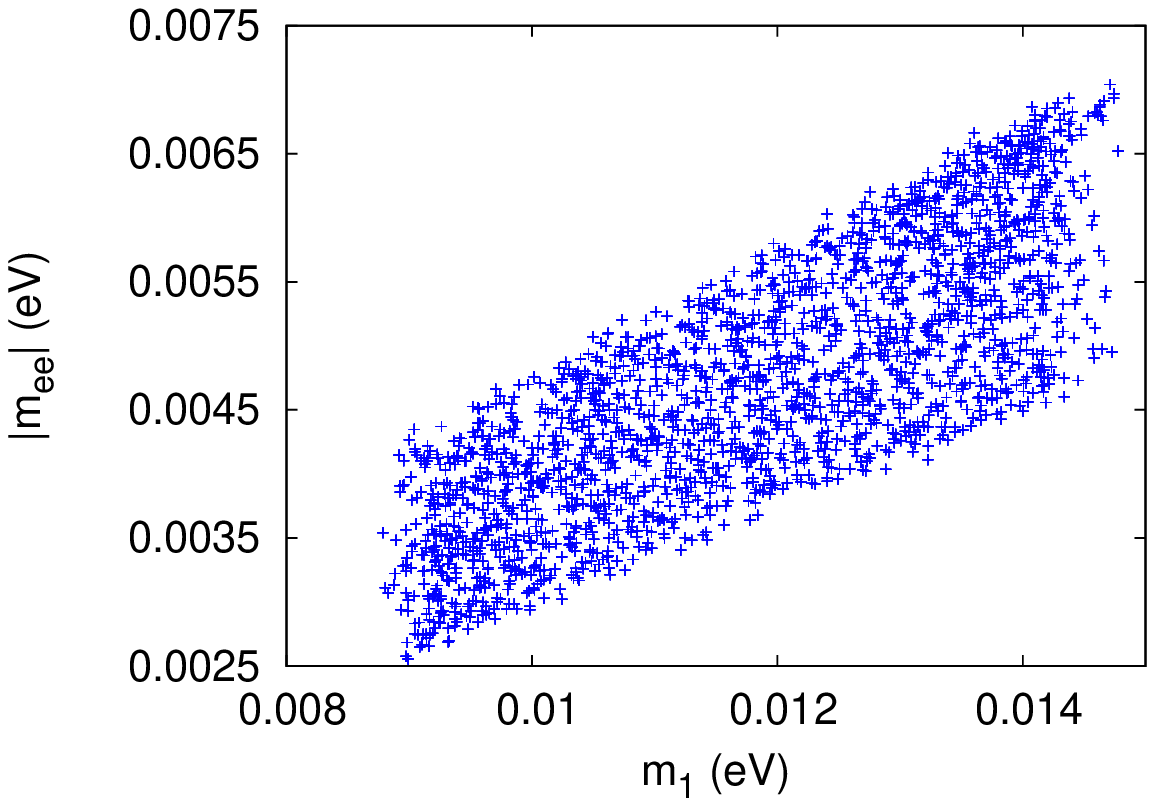, width=5.0cm, height=5.0cm}
\epsfig{file=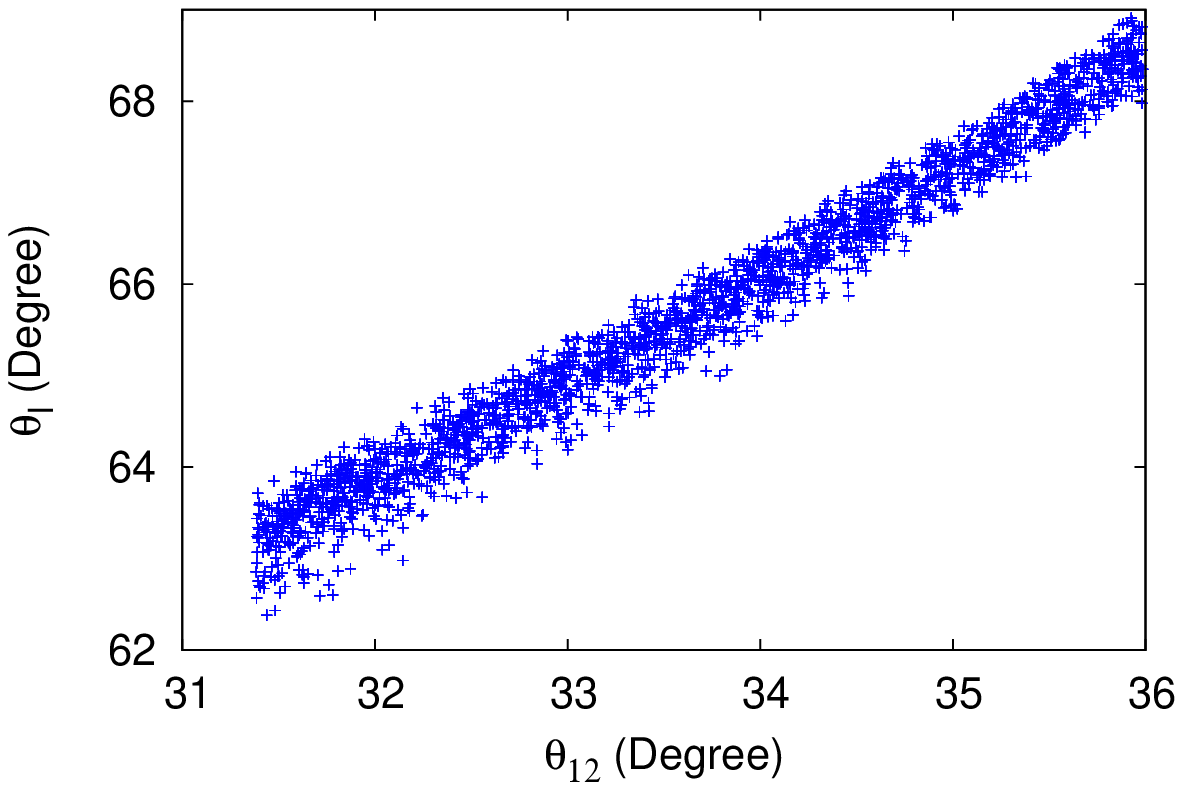, width=5.0cm, height=5.0cm}
\epsfig{file=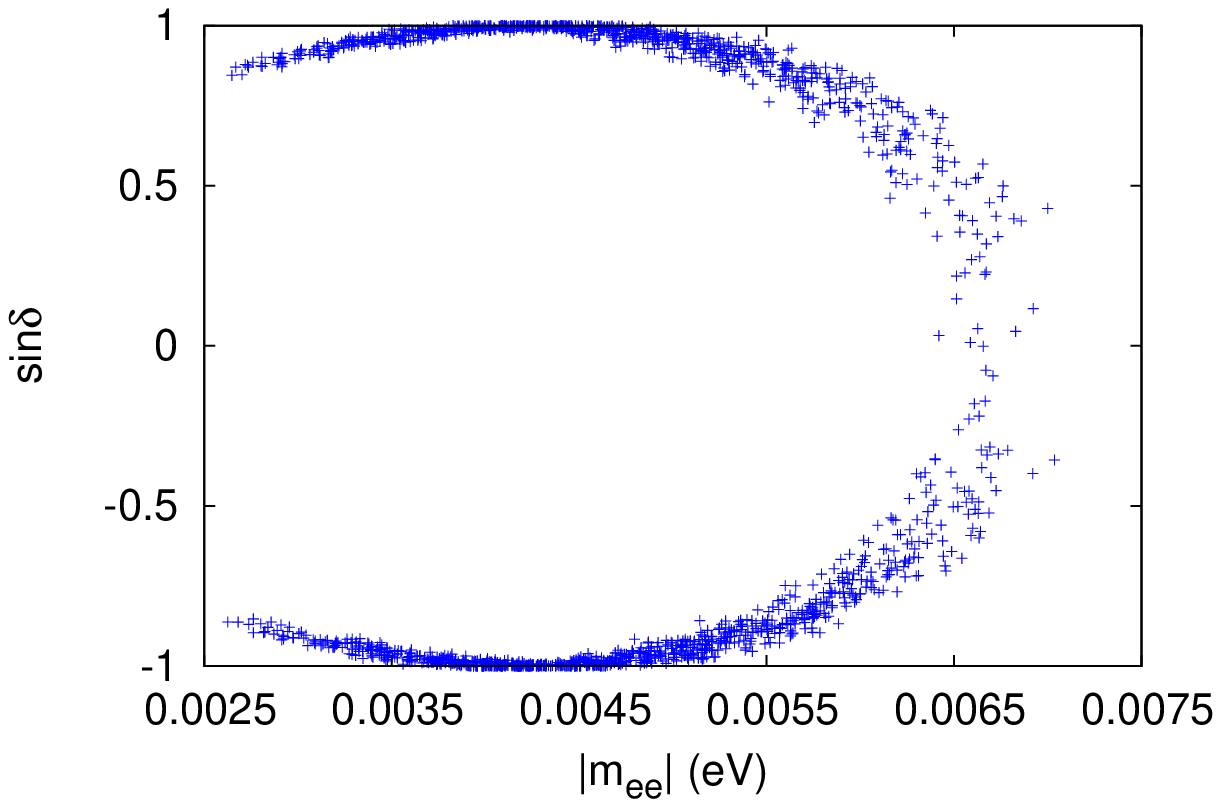, width=5.0cm, height=5.0cm}\\
\epsfig{file=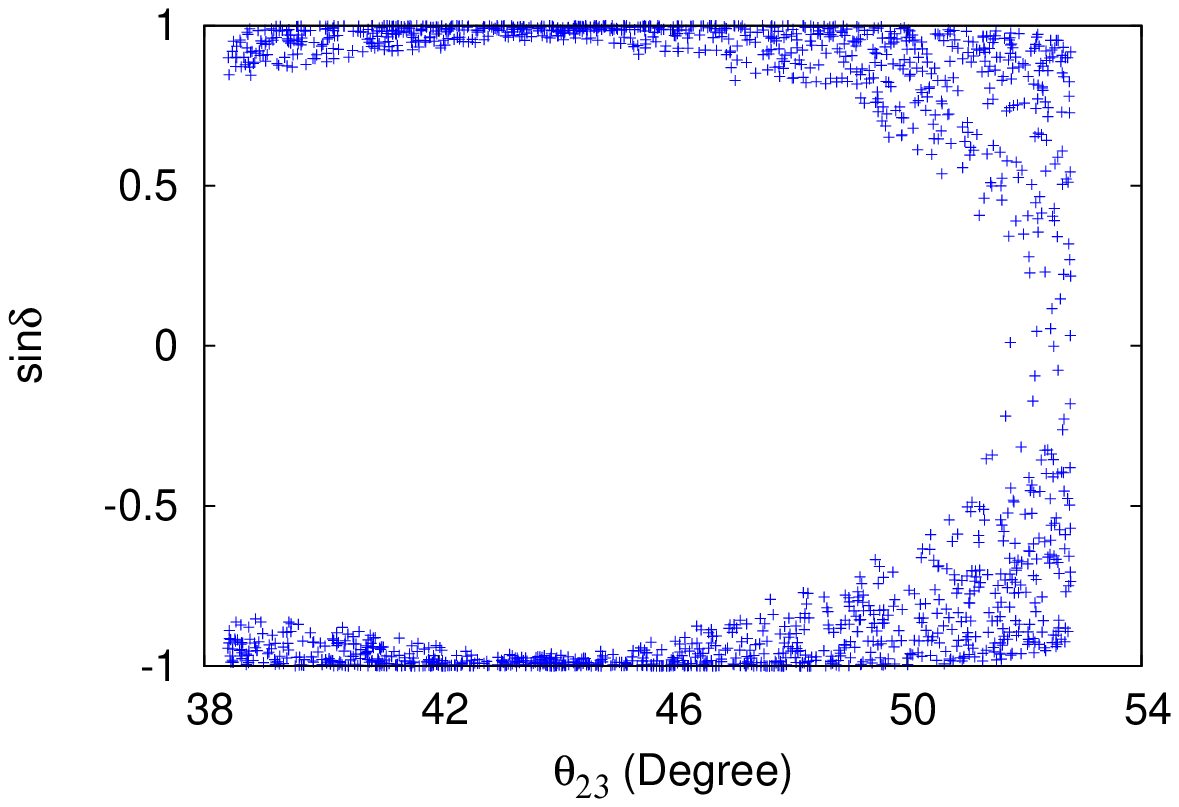, width=5.0cm, height=5.0cm}
\epsfig{file=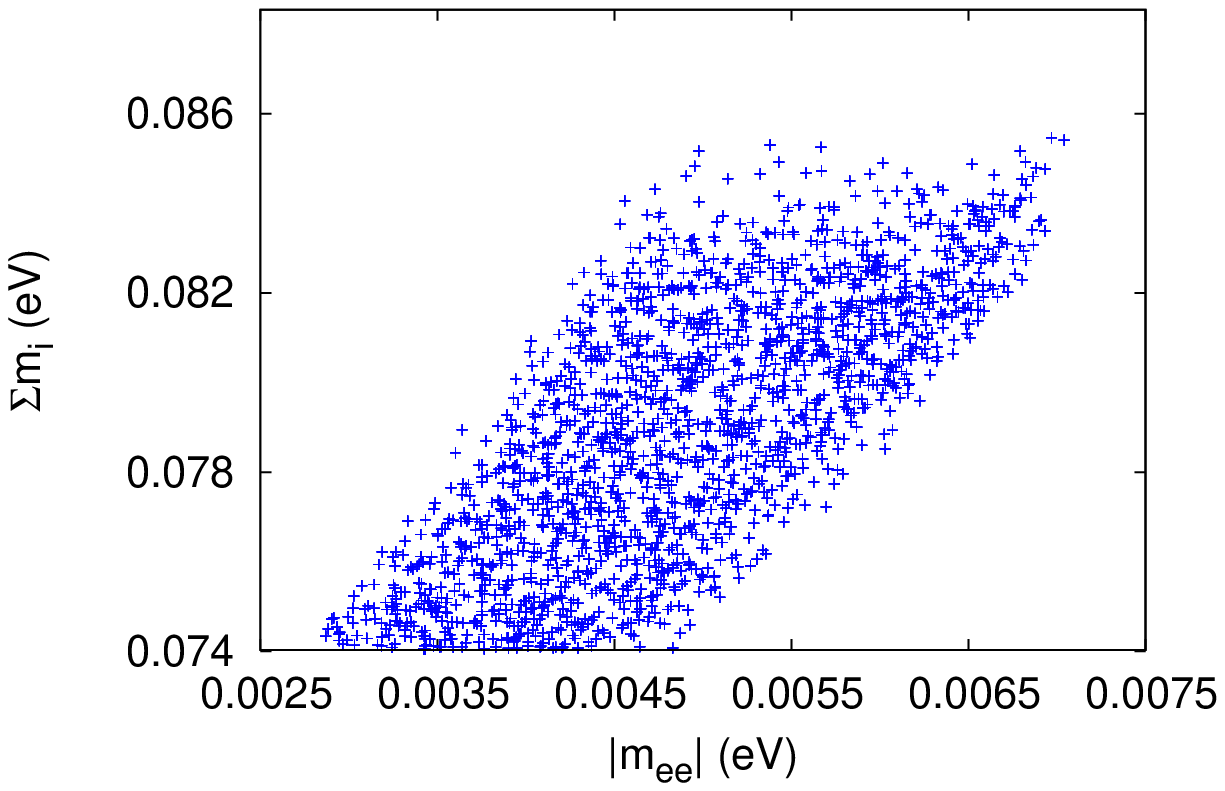, width=5.0cm, height=5.0cm}
\epsfig{file=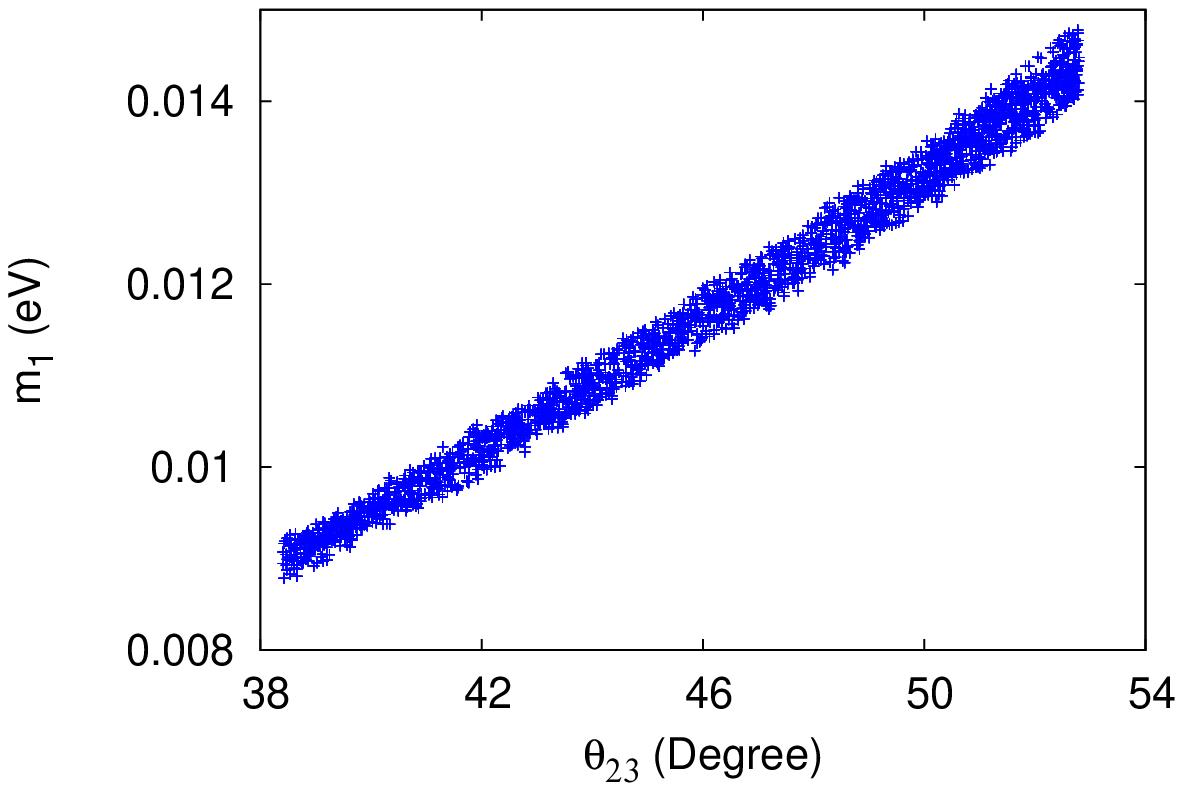, width=5.0cm, height=5.0cm}
\end{center}
\caption{Correlation plots between different parameters for NO in texture III-(H).}
\label{fig1}
\end{figure}

\begin{figure}[h]
\begin{center}
\epsfig{file=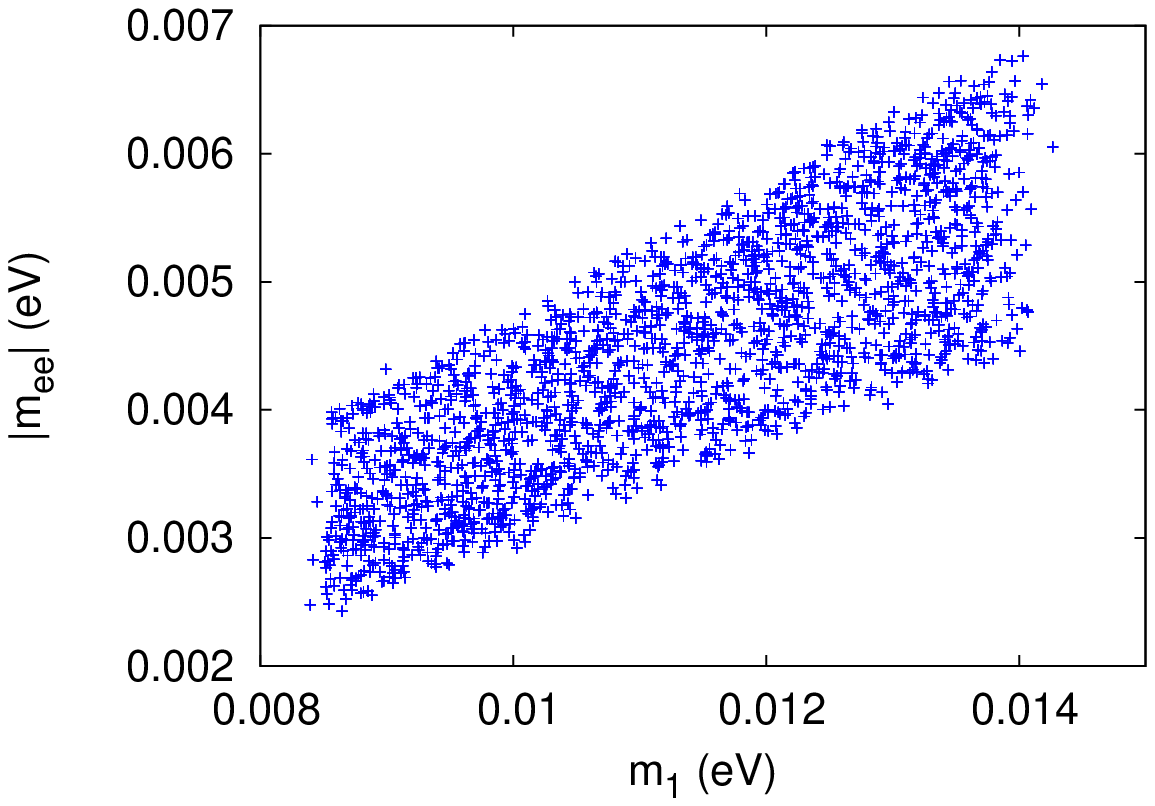, width=5.0cm, height=5.0cm}
\epsfig{file=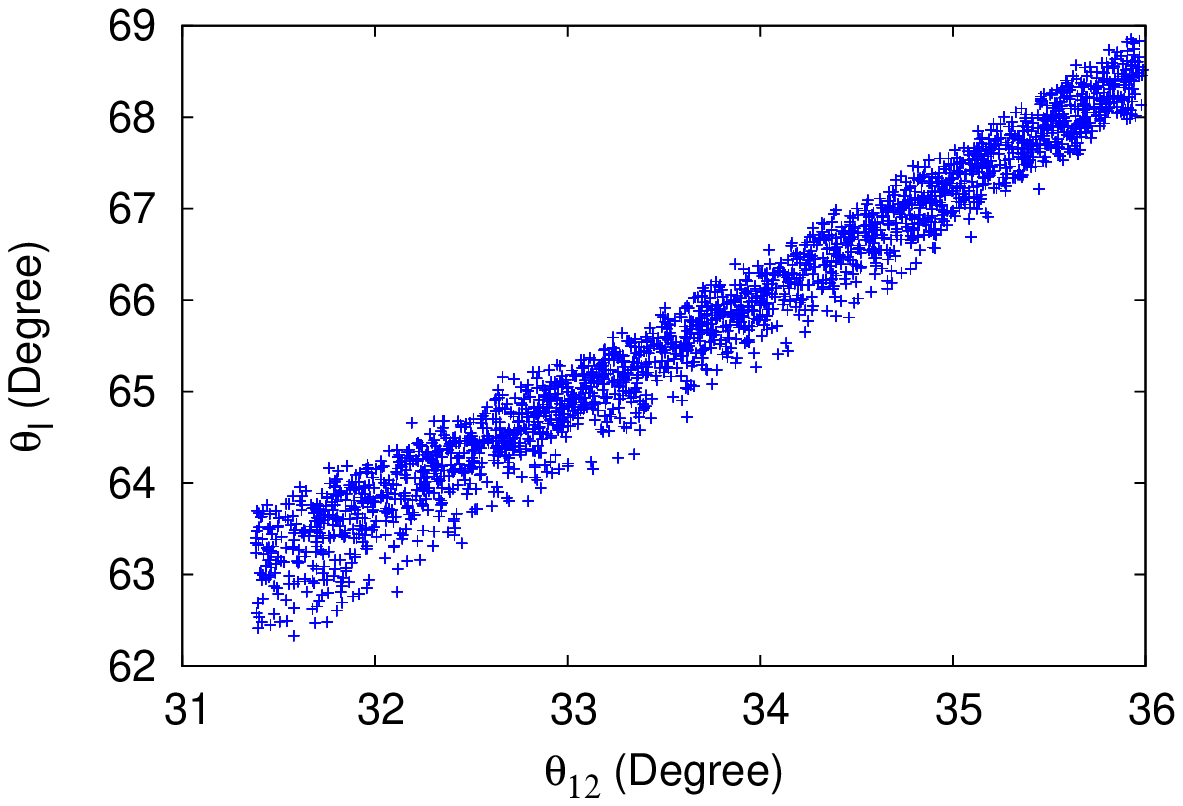, width=5.0cm, height=5.0cm}
\epsfig{file=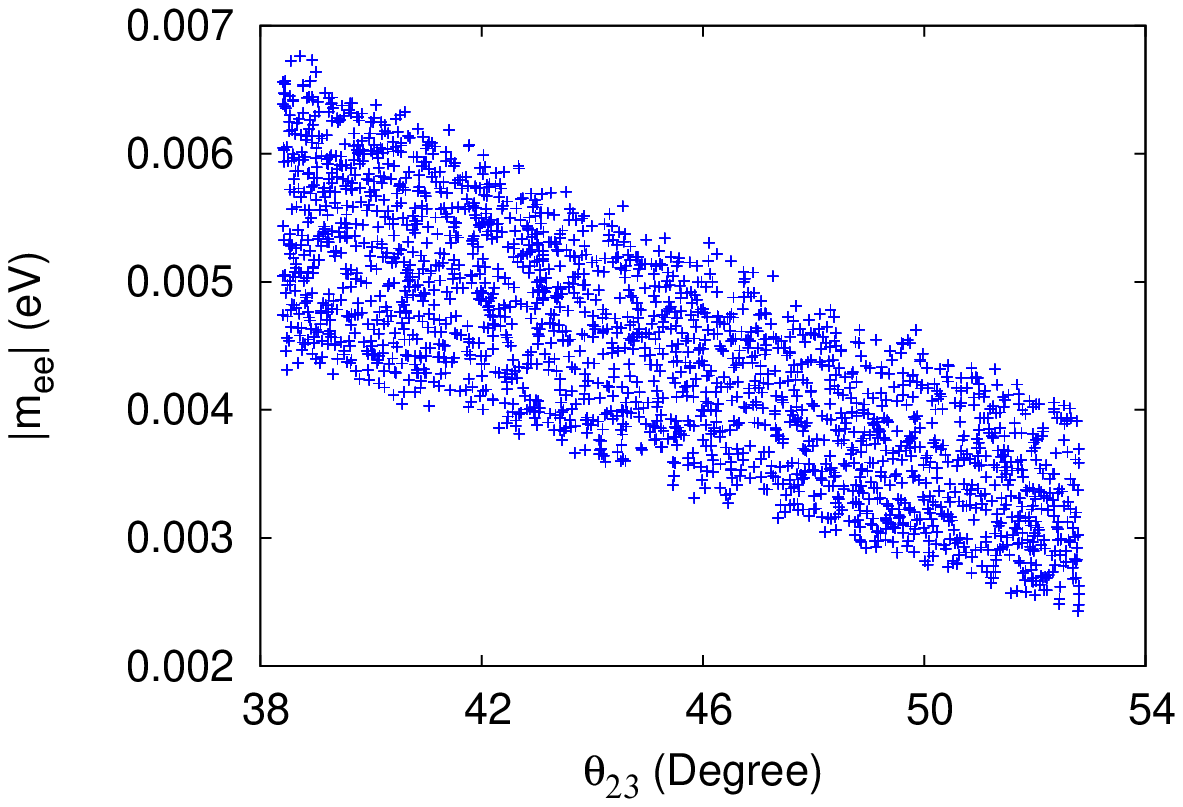, width=5.0cm, height=5.0cm}\\
\epsfig{file=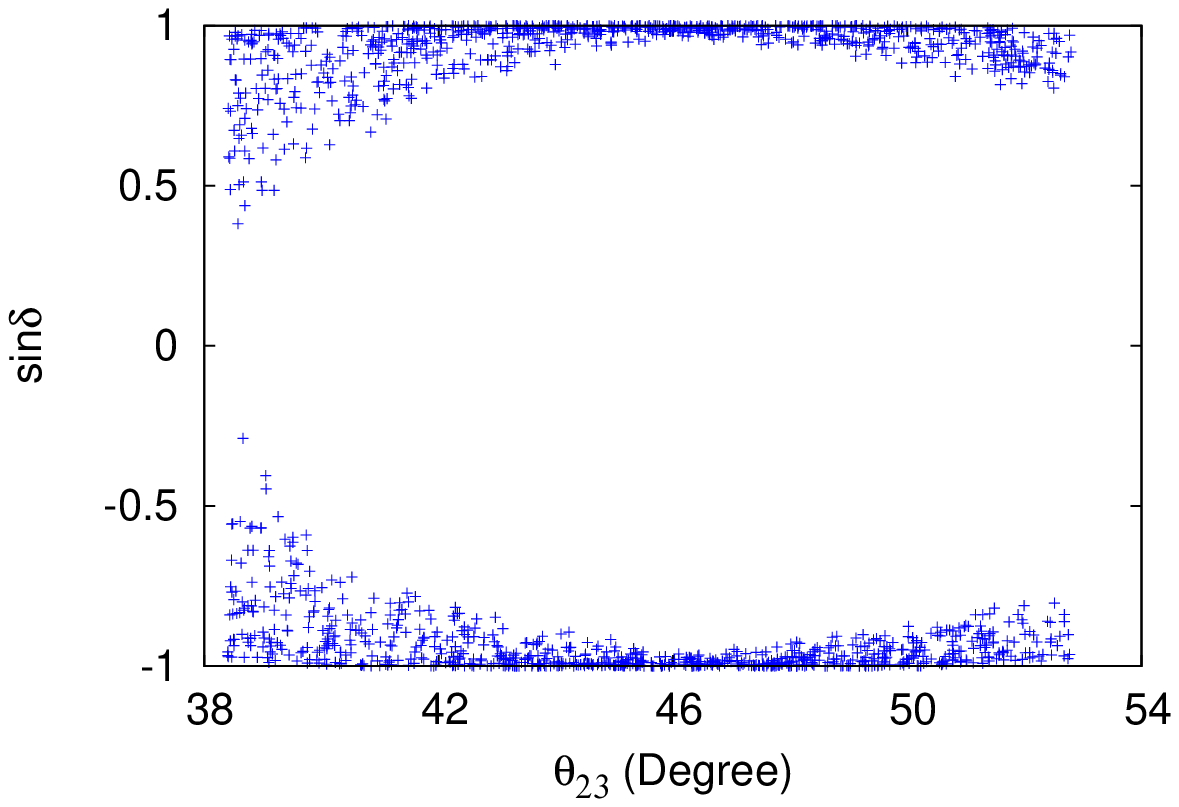, width=5.0cm, height=5.0cm}
\epsfig{file=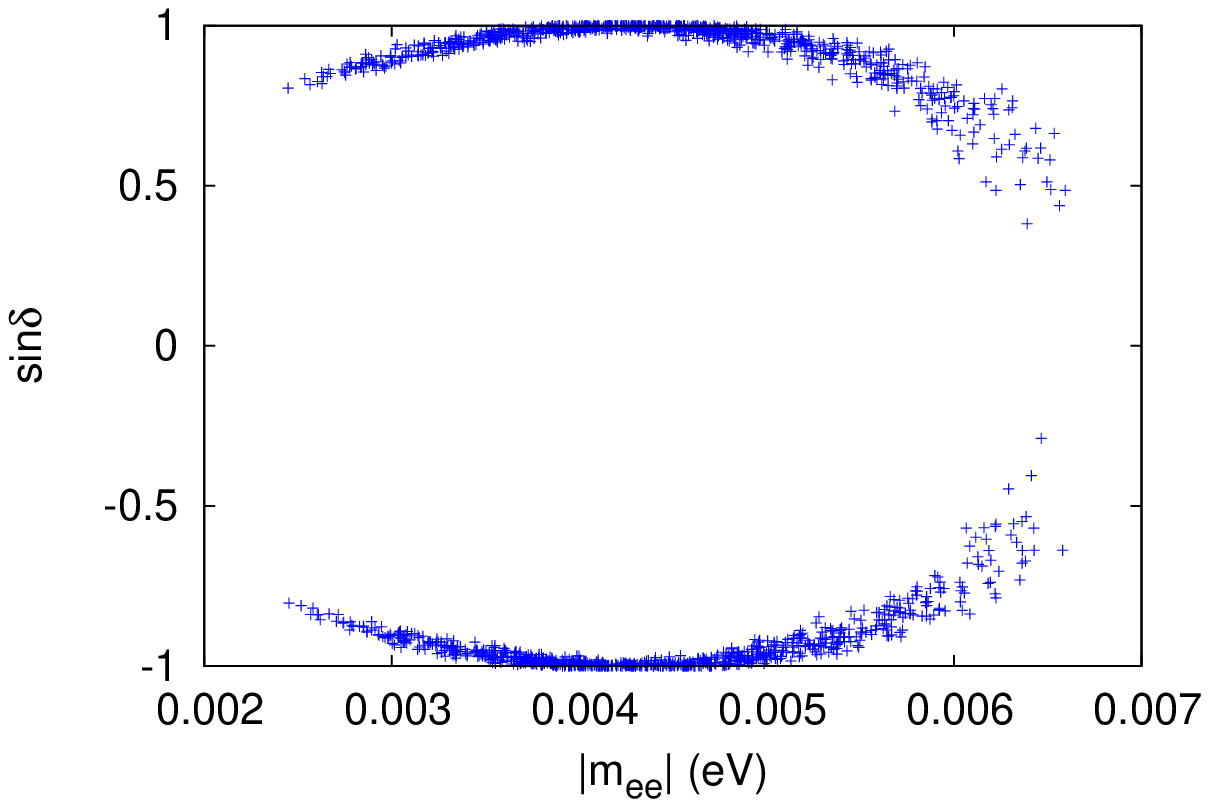, width=5.0cm, height=5.0cm}
\epsfig{file=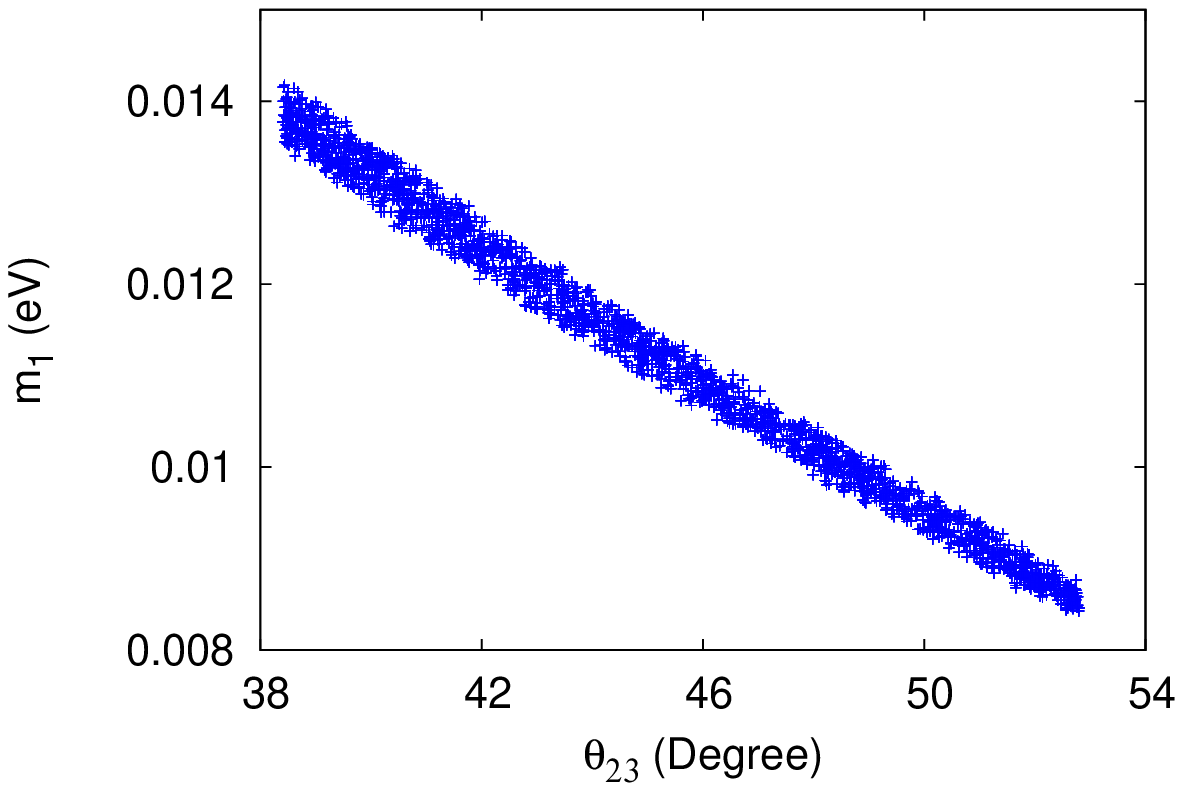, width=5.0cm, height=5.0cm}
\end{center}
\caption{Correlation plots between different parameters for NO in texture III-(I).}
\label{fig2}
\end{figure}

\textbf{Texture III-(H)}:
First, we analyze the texture structure III-(H) in which $M_{\nu}$ has vanishing cofactors corresponding to (1,3), (2,2) and (3,3) elements and $H_{l}$ has texture zeros at (1,3) and (2,3) positions. The texture structures of neutrino and charged lepton mass matrices have the following form:
\begin{eqnarray}
M_{\nu 8}=\left(
\begin{array}{ccc}
\times & \times & \Delta \\
\times & \Delta & \times \\
\Delta & \times & \Delta
\end{array}
\right)~~ \textrm{and} ~~
H_{l 3}=\left(
\begin{array}{ccc}
\times & \times & 0 \\
 \times & \times & 0 \\
 0 & 0 & m_{\tau}^2
\end{array}
\right).
\end{eqnarray}
All the non-zero elements of $M_{\nu}$ are, in general, complex. The neutrino mass matrix can be made real as $M_{\nu}=P_{\nu} M_{\nu}^{r} P_{\nu}^{T}$, with the phase matrix $P_{\nu} = $ diag$(e^{i \psi_{1}},e^{i \psi_{2}},e^{i \psi_{3}})$. The matrix $M_{\nu}^{r}$ is diagonalized by orthogonal matrix $O_{\nu}$ as
\begin{equation}
M_{\nu}^{r}=O_{\nu} M_{\nu}^{D} O_{\nu}^{T}
\end{equation} 
and the neutrino mixing matrix is $V_{\nu}=P_{\nu} O_{\nu}$. The PMNS mixing matrix is given by
\begin{equation}
U=V^{\dagger}_{l} V_{\nu}=V^{\dagger}_{l} P_{\nu} O_{\nu}.
\end{equation}
We use invariants Tr$[M_{\nu}^{r}]$, Tr$[M_{\nu}^{r^{2}}]$ and Det$[M_{\nu}^{r}]$ to redefine mass matrix elements in terms of mass eigenvalues. The eigenvalues of the neutrino mass matrix for NO are $m_1, -m_2$ and $m_3$. The eigenvalues $m_{2}$ and $m_{3}$ can be calculated from the mass-squared differences $\Delta m_{21}^{2}$ and $\Delta m_{31}^{2}$ using the relation
\begin{equation}
m_{2}=\sqrt{\Delta m_{21}^{2}+m_{1}^{2}} ~~\textrm{and}~~ m_{3}=\sqrt{\Delta m_{31}^{2}+m_{1}^{2}}.
\end{equation}\\
The orthogonal mixing matrix $O_{\nu}$ for NO is given by 
\begin{small}
\begin{equation}
O_{\nu 8| NO}=\left(
\begin{array}{ccc}
 -\frac{\sqrt{m_2 m_3} \sqrt{(m_2-m_1)(m_1+m_3)}}{\sqrt{(m_1+m_2)(m_3-m_1) a}} &
   -\frac{\sqrt{m_1 m_3} \sqrt{(m_2-m_1) (m_3-m_2)}}{\sqrt{(m_1+m_2)(m_2+m_3) a}} &
   \frac{\sqrt{m_1 m_2} \sqrt{(m_1+ m_3) (m_3-m_2)}}{\sqrt{(m_3-m_1)(m_2+m_3) a}} \\
 \frac{\sqrt{m_1(m_3-m_2)}}{\sqrt{(m_1+m_2) (m_3-m_1)}} & -\frac{\sqrt{m_2 (m_1+m_3)}}{\sqrt{(m_1+m_2)
   (m_2+m_3)}} & \frac{\sqrt{(m_1-m_2) m_3}}{\sqrt{(m_1-m_3) (m_2+m_3)}}  \\
 \frac{m_1 \sqrt{m_1 (m_3-m_2)}}{\sqrt{(m_1+m_2)(m_3-m_1) a}} & \frac{m_2 \sqrt{m_2 (m_1+m_3)}}{\sqrt{(m_1+m_2)
   (m_2+m_3) a}} & \frac{m_3 \sqrt{(m_2-m_1) m_3}}{\sqrt{(m_3-m_1) (m_2+m_3) a}}
\end{array}
\right)
\end{equation}
\end{small}
where $a=m_1 (m_2-m_3)+m_2 m_3$.
For IO, with neutrino mass eigenvalues ($-m_1, m_2, m_3$), the orthogonal matrix $O_{\nu}$ is given by 
\begin{equation}
O_{\nu 8|IO}=\left(
\begin{array}{ccc}
 -\frac{\sqrt{(m_2-m_1) (m_1-m_3)}
   \sqrt{m_2 m_3}}{\sqrt{b (m_1+m_2)(m_1+m_3)}} & \frac{\sqrt{m_1 m_3}
   \sqrt{(m_2-m_1) (m_2+m_3)}}{\sqrt{b(m_1+m_2) (m_2-m_3)}} &
   -\frac{\sqrt{m_1 m_2} \sqrt{(m_1-m_3)(m_2+m_3)}}{\sqrt{b (m_2-m_3)
   (m_1+m_3)}} \\
 -\frac{\sqrt{m_1 (m_2+m_3)}}{\sqrt{(m_1+m_2)(m_1+m_3)}} & \frac{\sqrt{m_2
   (m_1-m_3)}}{\sqrt{(m_1+m_2)(m_2-m_3)}} & \frac{\sqrt{(m_2-m_1) m_3}}{\sqrt{(m_2-m_3) (m_1+m_3)}}
   \\
 \frac{m_1 \sqrt{m_1 (m_2+m_3)}}{\sqrt{b(m_1+m_2) (m_1+m_3)}} & \frac{m_2
   \sqrt{m_2 (m_1-m_3)}}{\sqrt{b(m_1+m_2) (m_2-m_3)}} & \frac{m_3 \sqrt{(m_2-m_1) m_3}}{\sqrt{b
   (m_2-m_3) (m_1+m_3)}}
\end{array}
\right)
\end{equation}

where $b=m_1 (m_2+m_3)-m_2 m_3$. The mass eigenvalues can be calculated by using the relations
\begin{equation}
m_{2}=\sqrt{\Delta m_{23}^{2}+m_{3}^{2}} ~~\textrm{and}~~ m_{1}=\sqrt{\Delta m_{13}^{2}+m_{3}^{2}}~.
\end{equation}
The charged lepton mixing matrix is given by
\begin{equation}
V_{l3}=\left(
\begin{array}{ccc}
\cos \theta_{l} & e^{i \phi_l} \sin \theta_{l} & 0 \\
- e^{-i \phi_l} \sin \theta_{l} & \cos \theta_{l} & 0 \\
0 & 0 & 1
\end{array}
\right)
\end{equation}
where $\cos \theta_{l}=\sqrt{\frac{m_{\mu}^{2}-m}{m_{\mu}^{2}-m_{e}^{2}}}$ and $m_{e}^{2}<m<m_{\mu}^{2}$.\\
The lepton mixing matrix for texture structure III-(H) can be written as
\begin{eqnarray}
U&=&V_{l 3}^{\dagger}V_{\nu 8} \ \ 
= V_{l 3}^{\dagger} P_{\nu} O_{\nu 8|NO(IO)}\nonumber \\
&=& \left(
\begin{array}{ccc}
\cos \theta_{l} & -e^{i \phi_l} \sin \theta_{l} & 0 \\
e^{-i \phi_l} \sin \theta_{l} & \cos \theta_{l} & 0 \\
0 & 0 & 1
\end{array}
\right)  \left(
\begin{array}{ccc}
e^{i \psi_{1}} & 0 & 0 \\
0 & e^{i \psi_{2}} & 0 \\
0 & 0 & e^{i \psi_{3}}
\end{array}
\right) O_{\nu 8|NO(IO)}\nonumber \\
&=&P^{\prime} \left(
\begin{array}{ccc}
\cos \theta_{l} & -e^{i \eta} \sin \theta_{l} & 0 \\
\sin \theta_{l} & e^{i \eta} \cos \theta_{l} & 0 \\
0 & 0 & 1
\end{array}
\right) O_{\nu 8|NO(IO)}
\end{eqnarray}
where $\eta =\psi_2-\psi_1+\phi_l$ and the phase matrix $P^{\prime}=$ diag$(e^{i \psi_{1}},e^{i (\psi_{1}-\phi_l)},e^{i \psi_{3}})$.\\
\textbf{Texture III-(I)}:\\
For texture structure III-(I), $M_{\nu}$ has vanishing cofactors corresponding to (1,2), (2,2), and (3,3) elements while $H_{l}$ has zeros at (1,2) and (2,3) elements. The neutrino and the charged lepton mass matrices for texture III-(I) have the following form:
\begin{eqnarray}
M_{\nu 9}=\left(
\begin{array}{ccc}
\times & \Delta & \times \\
\Delta & \Delta & \times \\
\times & \times & \Delta
\end{array}
\right)~~\textrm{and}~~
H_{l2}=\left(
\begin{array}{ccc}
\times & 0 & \times \\
 0 & m_{\mu}^{2} & 0 \\
 \times & 0 & \times
\end{array}
\right).
\end{eqnarray}
The neutrino mass matrix $M_{\nu 9}$ is related to $M_{\nu 8}$ by permutation symmetry, $M_{\nu 9}=S_{23} M_{\nu 8} S_{23}^{T}$. Therefore, the neutrino mixing matrix for texture $M_{\nu 9}$ can be written in terms of $V_{\nu 8}$ as $V_{\nu 9}=S_{23} V_{\nu 8}$. Therefore, the PMNS mixing matrix for texture III-(I) is given by 
\begin{eqnarray}
U&=&V^{\dagger}_{l2} V_{\nu 9} = V^{\dagger}_{l2} S_{23} V_{\nu 8} = V^{\dagger}_{l2} S_{23} P_{\nu} O_{\nu 8}~\nonumber \\
 &=& \left(
\begin{array}{ccc}
\cos \theta_{l} & 0 & -e^{i \phi_l} \sin \theta_{l} \\
0 & 1 & 0 \\
e^{-i \phi_l} \sin \theta_{l} & 0 & \cos \theta_{l}
\end{array}
\right) 
\left(
\begin{array}{ccc}
1 & 0 & 0 \\
0 & 0 & 1 \\
0 & 1 & 0
\end{array}
\right)
 \left(
\begin{array}{ccc}
e^{i \psi_{1}} & 0 & 0 \\
0 & e^{i \psi_{2}} & 0 \\
0 & 0 & e^{i \psi_{3}}
\end{array}
\right) O_{\nu 8|NO(IO)}~ \nonumber \\
&=& P^{\prime} \left(
\begin{array}{ccc}
\cos \theta_{l} & -e^{i \eta} \sin \theta_{l} & 0 \\
0 & 0 & 1 \\
\sin \theta_{l} & e^{i \eta} \cos \theta_{l} & 0
\end{array}
\right) O_{\nu 8|NO(IO)}.
\end{eqnarray}
For NO and IO, the orthogonal matrix $O_{\nu 8}$ is given in Eqs. (22) and (23), respectively.\\
The three lepton mixing angles in terms of the lepton mixing matrix elements are given by 
\begin{eqnarray}
\sin \theta_{13}=|U_{13}|,~~ \sin \theta_{23}= \frac{|U_{23}|}{\sqrt{1-|U_{13}|^{2}}} ~~\textrm{and}~~ \sin \theta_{12}=\frac{|U_{12}|}{\sqrt{1-|U_{13}|^2}}.
\end{eqnarray}     
The CP violating phase $\delta$ can be calculated using Eq. (15):
\begin{equation}
\sin \delta = \frac{\textrm{Im}(U_{11} U_{22} U_{12}^\dagger U_{21}^\dagger)}{\sin\theta_{12} \sin\theta_{23} \sin\theta_{13} \cos\theta_{12} \cos\theta_{23} \cos^{2}\theta_{13}}
\end{equation}
where the elements of $U$ are given in Eq. (26) for texture III-(H) and Eq. (28) for texture III-(I). \\ 
Similarly, the three mixing angles can be calculated by substituting the elements of $U$ from Eqs.(26) and (28) into Eq.(29)  for textures III-(H) and III-(I), respectively. The effective Majorana mass $|m_{ee}|$ can be calculated using Eq. (16)\\
There exist following relations between neutrino oscillation parameters of texture structures III-(H) and III-(I):
\begin{equation}
\theta_{12}^{H}=\theta_{12}^{I}, \theta_{13}^{H}=\theta_{13}^{I}, \theta_{23}^{H}=\frac{\pi}{2}-\theta_{23}^{I}.
\end{equation}
\begin{table}[h]
\begin{center}
\begin{tabular}{|c|c|c|}
 \hline
Neutrino Parameter & Normal Ordering & Inverted Ordering \\
  & bfp $\pm 1 \sigma$ ~~ $3\sigma$ range  & bfp $\pm 1 \sigma$~~ $3\sigma$ range \\
 \hline 
$\theta_{12}^{\circ}$ & $33.56^{+0.77}_{-0.75}$ ~~ $31.38 \rightarrow 35.99$ & $33.56^{+0.77}_{-0.75}$ ~~ $31.38 \rightarrow 35.99$ \\
$\theta_{23}^{\circ}$ & $41.6^{+1.5}_{-1.2}$  ~~$38.4 \rightarrow 52.8$ & $50.0^{+1.1}_{-1.4}$ ~~ $38.8 \rightarrow 53.1$ \\
$\theta_{13}^{\circ}$ & $8.46^{+0.15}_{-0.15}$  ~~$7.99 \rightarrow 8.90$ & $8.49^{+0.15}_{-0.15}$ ~~ $8.03 \rightarrow 8.93$ \\
$\delta_{CP}^{\circ}$ & $261^{+51}_{-59}$ ~~ $0.0 \rightarrow 360$ & $277^{+ 40}_{-46}$ ~~ $145 \rightarrow 391$ \\
$\Delta m^{2}_{21}/10^{-5}$ eV$^2$ & $7.50^{+0.19}_{-0.17}$ ~~$7.03 \rightarrow 8.09$ & $7.50^{+0.19}_{-0.17}$ ~~ $7.03 \rightarrow 8.09$ \\
$\Delta m^{2}_{3 l}/10^{-3}$ eV$^2$ & $+2.524^{+0.039}_{-0.040}$ ~~ $2.407 \rightarrow 2.643$ & $-2.514^{+0.038}_{-0.041}$ ~~ $-2.635 \rightarrow -2.399$ \\
 \hline 
 \end{tabular}
\caption{Current neutrino oscillation parameters from global fits \cite{data} with $\Delta m^{2}_{3 l}\equiv \Delta m^{2}_{31}>0$ for NO and $\Delta m^{2}_{3 l}\equiv \Delta m^{2}_{32}<0$ for IO.}
\label{tab2}
\end{center}
\end{table}
For the numerical analysis, we have generated random numbers of the order of $\sim$ $10^7$ for parameters $\Delta m^{2}_{21}$ and $\Delta m^{2}_{31} (\Delta m^{2}_{32})$ for NO (IO) within their experimentally allowed $3 \sigma$ ranges. $m_{1} (m_{3})$ have been generated randomly between $0$ - $0.33$ eV. We vary the parameter $\eta$ randomly within the range (0-2$\pi$) and parameter $m$ has been varied randomly with in the range $(m_{e}^2$ - $m_{\mu}^2)$ for texture III-(H) and $(m_{e}^2$ - $m_{\tau}^2)$ for texture III-(I). We use the experimental constraints on neutrino oscillation parameters as given in Table \ref{tab2}. In this analysis, the upper bound on sum of neutrino masses is set to be $\sum m_{i}\leq 1$ eV. It turns out that textures III-(H) and III-(I) are, phenomenologically, viable only for normal mass ordering. Figs. \ref{fig1} and \ref{fig2} depict the predictions for textures III-(H) and III-(I), respectively.\\
It can be seen from Fig. \ref{fig1} that $\sin\delta$ lies in the range ($-$1 - 1) and the Jarlskog CP invariant parameter $J_{CP}$ varies in the range ($-$0.036 - 0.036) for NO in III-(H) whereas for texture III-(I) the ranges for $\sin\delta$ and $J_{CP}$ as shown in Fig. \ref{fig2} are ($-$1 - 1) and $\pm(0.004$ - $0.036)$, respectively. The correlation plots in $(\theta_{23}$-$\sin \delta)$ plane in Figs. \ref{fig1} and \ref{fig2} show that $\delta$ is more favored to lie near $\delta \sim \pm \frac{\pi}{2}$. These results are consistent with the recent observations in the long baseline neutrino oscillation experiments like T2K and NOvA \cite{t2knova} which show a preference for the CP violating phase $\delta$ to lie around $\delta \sim -\frac{\pi}{2}$. The range for smallest neutrino mass $m_{1}$ is (0.0087 - 0.015) eV for texture III-(H) and (0.0084 - 0.014) eV for texture III-(I). The sum of neutrino masses $\sum m_{i}$ lies in the ranges (0.071 - 0.086) eV and (0.07 - 0.085) eV for textures III-(H) and III-(I), respectively. The charged lepton correction $\theta_{l}$ for Class-III turns out to be very large and lies in the range ($62^{\circ}$ - $69^{\circ}$) for both allowed textures. Parameter $|m_{ee}|$ is constrained to lie in the range (0.0025 - 0.0071) eV for texture III-(H) and (0.0024 - 0.007) eV for III-(I). All textures of Class-III with inverted mass ordering are ruled out at $3\sigma$ CL.\\


\subsection*{Class-IV}
The textures IV-(A), IV-(B) and IV-(C) are, phenomenologically, non-viable for the current $3\sigma$ ranges of neutrino oscillation parameters due to the presence of one zero element in the lepton mixing matrix $U$. We discuss the viable textures of Class-IV and their phenomenology below.

\textbf{Texture IV-(D)} Here $M_{\nu}$ has vanishing cofactors corresponding to elements (1,1), (1,2), and (1,3) which is equivalent to scaling neutrino mass matrix \cite{dev2} and charged lepton mass matrix $H_{l}$ has zeros at (1,3) and (2,3) positions. The mass matrices $M_{\nu}$ and $H_{l}$ are given by
\begin{eqnarray}
M_{\nu 14}&=&\left(
\begin{array}{ccc}
\Delta & \Delta & \Delta \\
\Delta & \times & \times \\
\Delta & \times & \times
\end{array}
\right) \equiv
\left(
\begin{array}{ccc}
 A e^{i \phi_a } & B e^{i \phi_b } & \frac{B e^{i \phi_b }}{c} \\
 B e^{i \phi_b } & D e^{i \phi_d } & \frac{D e^{i \phi_d}}{c} \\
 \frac{B e^{i \phi_b }}{c} & \frac{D e^{i \phi_d}}{c} & \frac{D e^{i \phi_d}}{c^2}
\end{array}
\right),\\
H_{l3}&=&\left(
\begin{array}{ccc}
\times & \times & 0 \\
 \times & \times & 0 \\
 0 & 0 & m_{\tau}^{2}
\end{array}
\right).
\end{eqnarray}
The neutrino mass matrix which, in general, is complex symmetric for this class can not be diagonalized directly due to the presence of non-removable phase. Instead, we diagonalize the Hermitian product $M_{\nu} M_{\nu}^{\dagger}$ which gives neutrino mixing matrix as
\begin{equation}
V_{\nu}^{\dagger} M_{\nu 14} M^{\dagger}_{\nu 14} V_{\nu} = ~\textrm{diag}(m_{1}^{2},m_{2}^{2},m_{3}^{2}).
\end{equation} 
Therefore, we have following Hermitian matrix
\begin{eqnarray}
M_{\nu 14} M^{\dagger}_{\nu 14}=\left(
\begin{array}{ccc}
 a & b e^{i \phi } & \frac{b e^{i \phi }}{c} \\
 b e^{-i \phi } & d & \frac{d}{c} \\
 \frac{b e^{-i \phi }}{c} & \frac{d}{c} & \frac{d}{c^2}
\end{array}
\right)= P^{\dagger}_{\nu} \left(
\begin{array}{ccc}
 a & b & \frac{b}{c} \\
 b & d & \frac{d}{c} \\
 \frac{b}{c} & \frac{d}{c} & \frac{d}{c^2}
\end{array}
\right)
P_{\nu}
\end{eqnarray}
where 
\begin{eqnarray}
a &= & A^2+B^2(1+\frac{1}{c^2}),\nonumber\\
b e^{i \phi } &=& A B e^{-i (\phi_a + \phi_b)}+B D e^{i (\phi_b + \phi_d)}(1+\frac{1}{c^2}),\\
d &=& B^2+D^2(1+\frac{1}{c^2}) \nonumber
\end{eqnarray}
and $P_{\nu}=$ diag$(e^{i \phi },1,1)$ is the phase matrix. The mixing matrix for $M_{\nu 14}$ is given by
\begin{eqnarray}
V_{\nu 14}&=&P_{\nu} O_{\nu 14}
\end{eqnarray}
where $O_{\nu 14}$ is an orthogonal unitary matrix which diagonalizes the real symmetric mass matrix given in Eq.(35).\\
For NO, the neutrino mass eigenvalues are $m_{1}=0$, $m_{2}=\sqrt{\Delta m^{2}_{21}}$, $m_{3}=\sqrt{\Delta m^{2}_{31}}$ and the orthogonal matrix $O_{\nu 14}$ is given by
\begin{eqnarray}
O_{\nu 14|NO}=\left(
\begin{array}{ccc}
 0 & \frac{a c^2-\left(c^2+1\right) d-\sqrt{x}}{\sqrt{4 b^2 c^4+4
   b^2 c^2+\left(-a c^2+d c^2+d+\sqrt{x}\right)^2}} & \frac{a
   c^2-\left(c^2+1\right) d+\sqrt{x}}{\sqrt{4 b^2 c^4+4 b^2
   c^2+\left(a c^2-\left(c^2+1\right) d+\sqrt{x}\right)^2}} \\
 -\frac{1}{\sqrt{c^2+1}} & \frac{2 b c^2}{\sqrt{4 b^2 c^4+4 b^2
   c^2+\left(-a c^2+d c^2+d+\sqrt{x}\right)^2}} & \frac{2 b
   c^2}{\sqrt{4 b^2 c^4+4 b^2 c^2+\left(a c^2-\left(c^2+1\right)
   d+\sqrt{x}\right)^2}} \\
 \frac{c}{\sqrt{c^2+1}} & \frac{2 b c}{\sqrt{4 b^2 c^4+4 b^2
   c^2+\left(-a c^2+d c^2+d+\sqrt{x}\right)^2}} & \frac{2 b
   c}{\sqrt{4 b^2 c^4+4 b^2 c^2+\left(a c^2-\left(c^2+1\right)
   d+\sqrt{x}\right)^2}}
\end{array}
\right)
\end{eqnarray}
where 
\begin{eqnarray}
x&=&\left(c^2 (d-a)+d\right)^2+4 b^2 \left(c^4+c^2\right),\nonumber \\
c&=&\frac{\sqrt{d}}{\sqrt{-a-d+m_{2}^{2}+m_{3}^{2}}}, \\
\textrm{and} ~ b&=&\frac{\sqrt{d} \sqrt{(m_{3}^{2}-a)\nonumber
   (a-m_{2}^{2})}}{\sqrt{-a+m_{2}^{2}+m_{3}^{2}}}.
\end{eqnarray}
The free parameters $a,d$ are constrained to lie in the range $m_{2}^{2}<a<m_{3}^{2}$ and $d<m_{3}^{2}$.
For IO, the mass eigenvalues are $m_{1}=\sqrt{\Delta m^{2}_{23}-\Delta m^{2}_{21}}, m_{2}=\sqrt{\Delta m^{2}_{23}}$ and $m_{3}=0$ and the corresponding orthogonal matrix $O_{\nu 14}$ is given by
\begin{eqnarray}
O_{\nu 14|IO}=\left(
\begin{array}{ccc}
 \frac{a c^2-\left(c^2+1\right) d-\sqrt{x}}{\sqrt{4 b^2 c^4+4 b^2
   c^2+\left(-a c^2+d c^2+d+\sqrt{x}\right)^2}} & \frac{a
   c^2-\left(c^2+1\right) d+\sqrt{x}}{\sqrt{4 b^2 c^4+4 b^2
   c^2+\left(a c^2-\left(c^2+1\right) d+\sqrt{x}\right)^2}} & 0 \\
 \frac{2 b c^2}{\sqrt{4 b^2 c^4+4 b^2 c^2+\left(-a c^2+d
   c^2+d+\sqrt{x}\right)^2}} & \frac{2 b c^2}{\sqrt{4 b^2 c^4+4
   b^2 c^2+\left(a c^2-\left(c^2+1\right) d+\sqrt{x}\right)^2}} &
   -\frac{1}{\sqrt{c^2+1}} \\
 \frac{2 b c}{\sqrt{4 b^2 c^4+4 b^2 c^2+\left(-a c^2+d
   c^2+d+\sqrt{x}\right)^2}} & \frac{2 b c}{\sqrt{4 b^2 c^4+4 b^2
   c^2+\left(a c^2-\left(c^2+1\right) d+\sqrt{x}\right)^2}} &
   \frac{c}{\sqrt{c^2+1}}
\end{array}
\right)
\end{eqnarray}
where
\begin{eqnarray}
x&=&\left(c^2 (d-a)+d\right)^2+4 b^2 \left(c^4+c^2\right),\nonumber \\
c&=&\frac{\sqrt{d}}{\sqrt{-a-d+m_{1}^{2}+m_{2}^{2}}},\\
b&=&\frac{\sqrt{d} \sqrt{(m_{2}^{2}-a)
   (a-m_{1}^{2})}}{\sqrt{-a+m_{1}^{2}+m_{2}^{2}}} \nonumber
\end{eqnarray}
with $m_{1}^{2}<a<m_{2}^{2}$ and $d<m_{2}^{2}$.\\
The PMNS mixing matrix for texture IV-(D) is given by
\begin{eqnarray}
U&=&V_{l3}^{\dagger} V_{\nu 14}\nonumber\\
&=&V_{l 3}^{\dagger} P_{\nu} O_{\nu 14|NO(IO)}\nonumber \\
&=& \left(
\begin{array}{ccc}
\cos \theta_{l} & -e^{i \phi_l} \sin \theta_{l} & 0 \\
e^{-i \phi_l} \sin \theta_{l} & \cos \theta_{l} & 0 \\
0 & 0 & 1
\end{array}
\right)  \left(
\begin{array}{ccc}
e^{i \phi} & 0 & 0 \\
0 & 1 & 0 \\
0 & 0 & 1
\end{array}
\right) O_{\nu 14|NO(IO)}\nonumber \\
&=& \left(
\begin{array}{ccc}
e^{i \phi} & 0 & 0 \\
0 & e^{i (\phi-\phi_l)} & 0 \\
0 & 0 & 1
\end{array}
\right) \left(
\begin{array}{ccc}
\cos \theta_{l} & -e^{i (\phi_l-\phi)} \sin \theta_{l} & 0 \\
\sin \theta_{l} & e^{i (\phi_l-\phi)} \cos \theta_{l} & 0 \\
0 & 0 & 1
\end{array}
\right) O_{\nu 14|NO(IO)}\nonumber \\
&=&P^{\prime} \left(
\begin{array}{ccc}
\cos \theta_{l} & -e^{i \eta} \sin \theta_{l} & 0 \\
\sin \theta_{l} & e^{i \eta} \cos \theta_{l} & 0 \\
0 & 0 & 1
\end{array}
\right) O_{\nu 14|NO(IO)}
\end{eqnarray}
where $\eta =\phi_l-\phi$ and $\cos \theta_{l}=\sqrt{\frac{m^{2}_{\mu}-m}{m^{2}_{\mu}-m^{2}_{e}}}$ for $m^{2}_{e}<m<m^{2}_{\mu}$. The phase matrix is given by $P^{\prime}=$diag$(e^{i \psi},e^{i (\phi-\phi_l)},1)$.
For IO of texture IV-(D), neutrino oscillation parameters are related as
\begin{equation}
\sin\theta_{13}=\frac{\sin \theta_{l}}{\sqrt{c^{2}+1}},~~ \tan\theta_{23}=\frac{\cos \theta_{l}}{c}~.
\end{equation}

\textbf{Texture IV-(E)}\\
Texture IV-(E) has three vanishing cofactors in $M_{\nu}$ corresponding to elements (1,1), (1,2), (1,3) and 
two zeros in $H_{l}$ at (1,2) and (2,3) positions. The mass matrices of neutrinos and charged leptons are given by
\begin{eqnarray}
M_{\nu 14}=\left(
\begin{array}{ccc}
\Delta & \Delta & \Delta \\
\Delta & \times & \times \\
\Delta & \times & \times
\end{array}
\right)~~\textrm{and}~~
H_{l2}=\left(
\begin{array}{ccc}
\times & 0 & \times \\
 0 & m_{\mu}^{2} & 0 \\
 \times & 0 & \times
\end{array}
\right).
\end{eqnarray}

The charged lepton mixing matrix for $H_{l2}$ is given by
\begin{equation}
V_{l2}=\left(
\begin{array}{ccc}
 \cos \theta_{l} & 0 & e^{i \phi_l} \sin \theta_{l} \\
 0 & 1 & 0 \\
 -e^{-i \phi_l} \sin \theta_{l} & 0 & \cos \theta_{l}
\end{array}
\right)
\end{equation}
where $\cos \theta_{l}=\sqrt{\frac{m_{\tau}^{2}-m}{m_{\tau}^{2}-m_{e}^{2}}}$ for $m_{e}^{2}<m<m_{\tau}^{2}$.\\

The PMNS mixing matrix for this texture is given by
\begin{eqnarray}
U&=&V_{l2}^{\dagger} V_{\nu 14},\nonumber \\
&=& V_{l2}^{\dagger} P_{\nu} O_{\nu 14|NO(IO)}, \nonumber \\
&=& P^{\prime} \left(
\begin{array}{ccc}
 \cos \theta_{l} & 0 & -e^{i \eta} \sin \theta_{l} \\
 0 & 1 & 0 \\
 \sin \theta_{l} & 0 & e^{i \eta} \cos \theta_{l}
\end{array}
\right) O_{\nu 14|NO(IO)}
\end{eqnarray}
where $\eta =\phi_l-\phi$ and $P^{\prime}=$ diag$(e^{i \psi},e^{i (\phi-\phi_l)},1)$. The orthogonal matrix $O_{\nu 14|NO(IO)}$ is given in Eqs.(38) and (40) for normal and inverted mass orderings.
Neutrino oscillation parameters for IO of texture IV-(E) are related as follows:
\begin{equation}
\sin\theta_{13}=\sin \theta_{l} \frac{c}{\sqrt{c^{2}+1}},~~ \tan\theta_{23}=\frac{\sec \theta_{l}}{c}.
\end{equation}

\begin{figure}[h]
\begin{center}
\epsfig{file=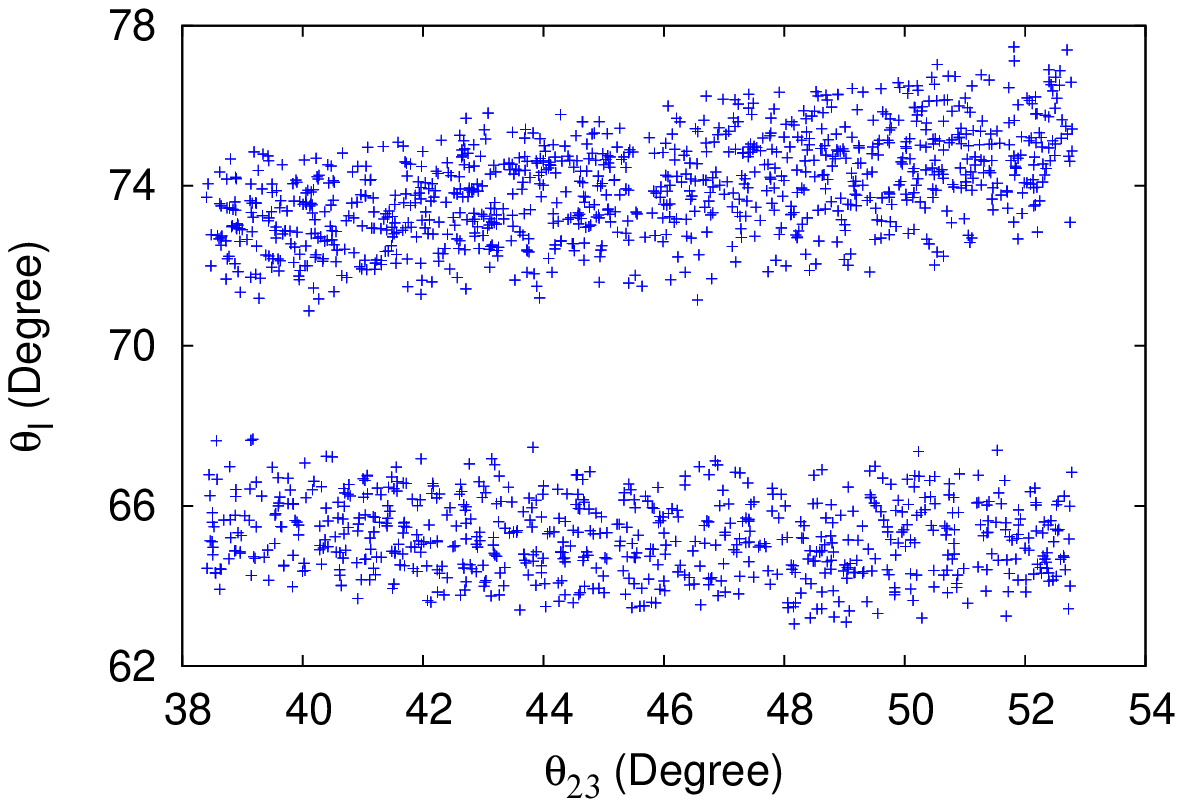, width=5.0cm, height=5.0cm}
\epsfig{file=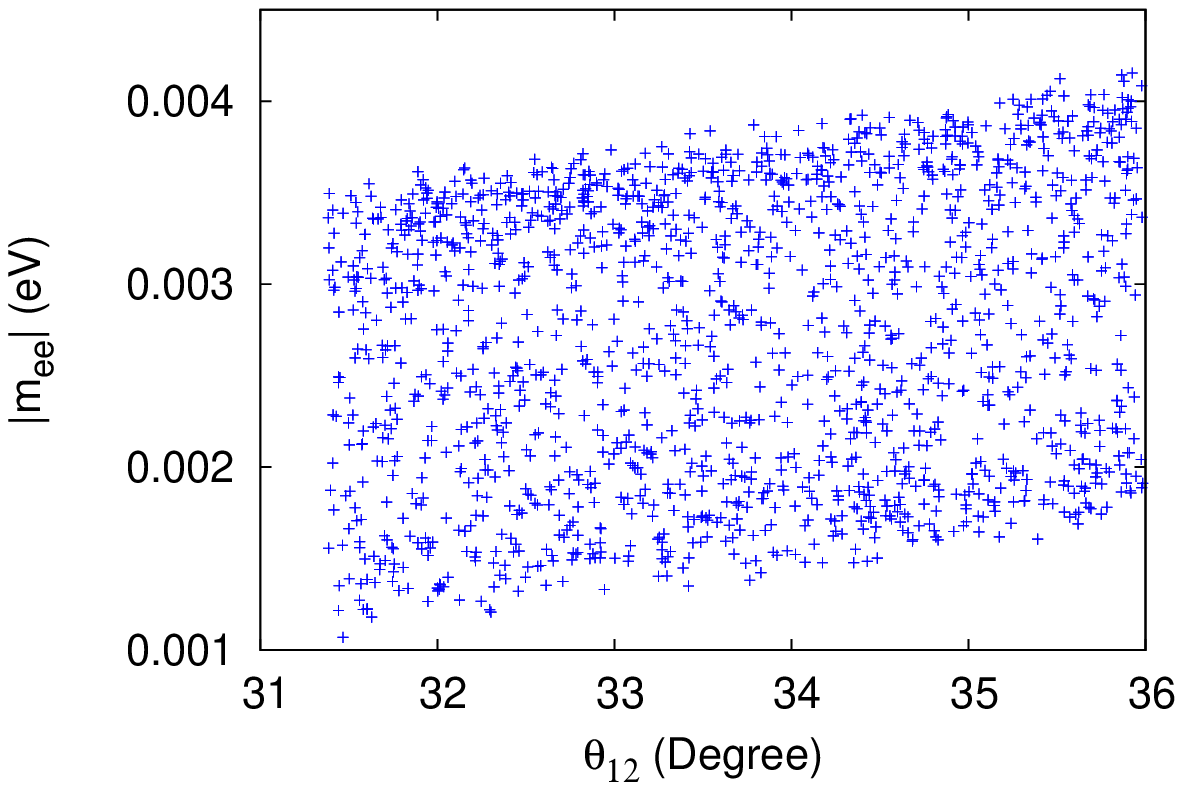, width=5.0cm, height=5.0cm}
\epsfig{file=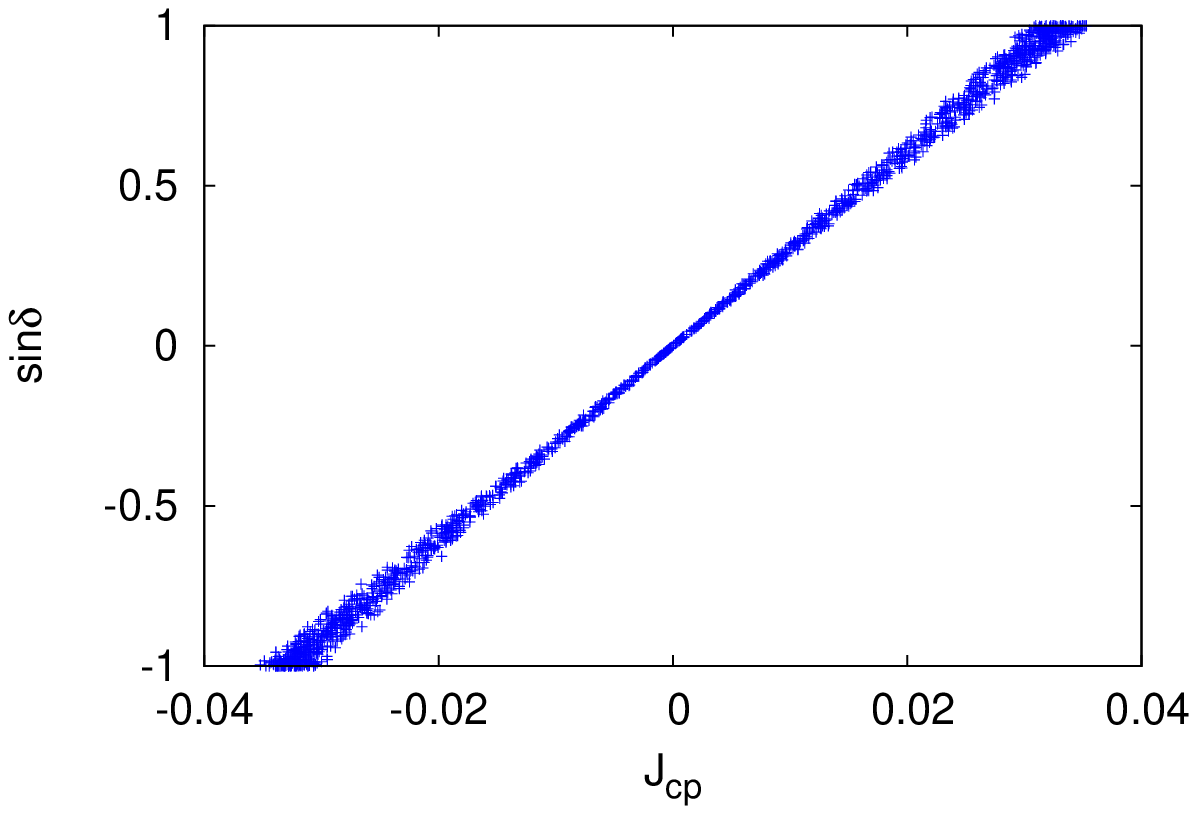, width=5.0cm, height=5.0cm}\\
\epsfig{file=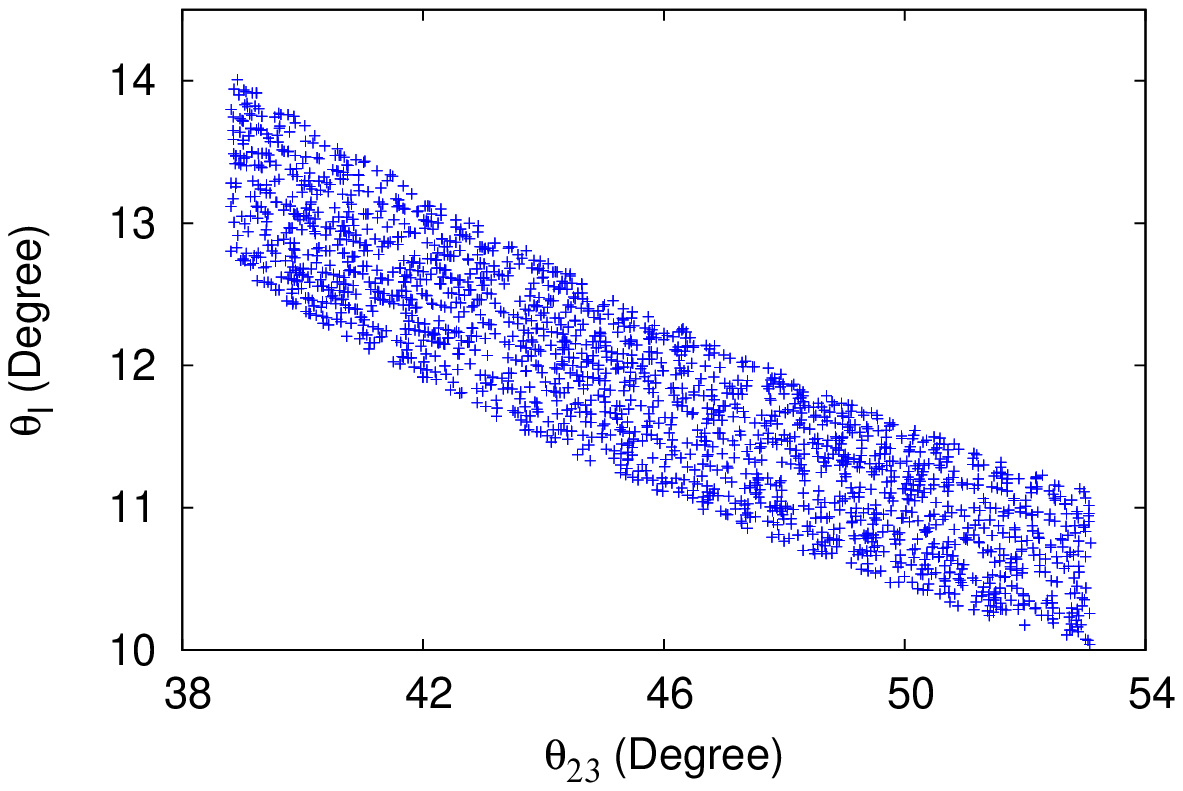, width=5.0cm, height=5.0cm}
\epsfig{file=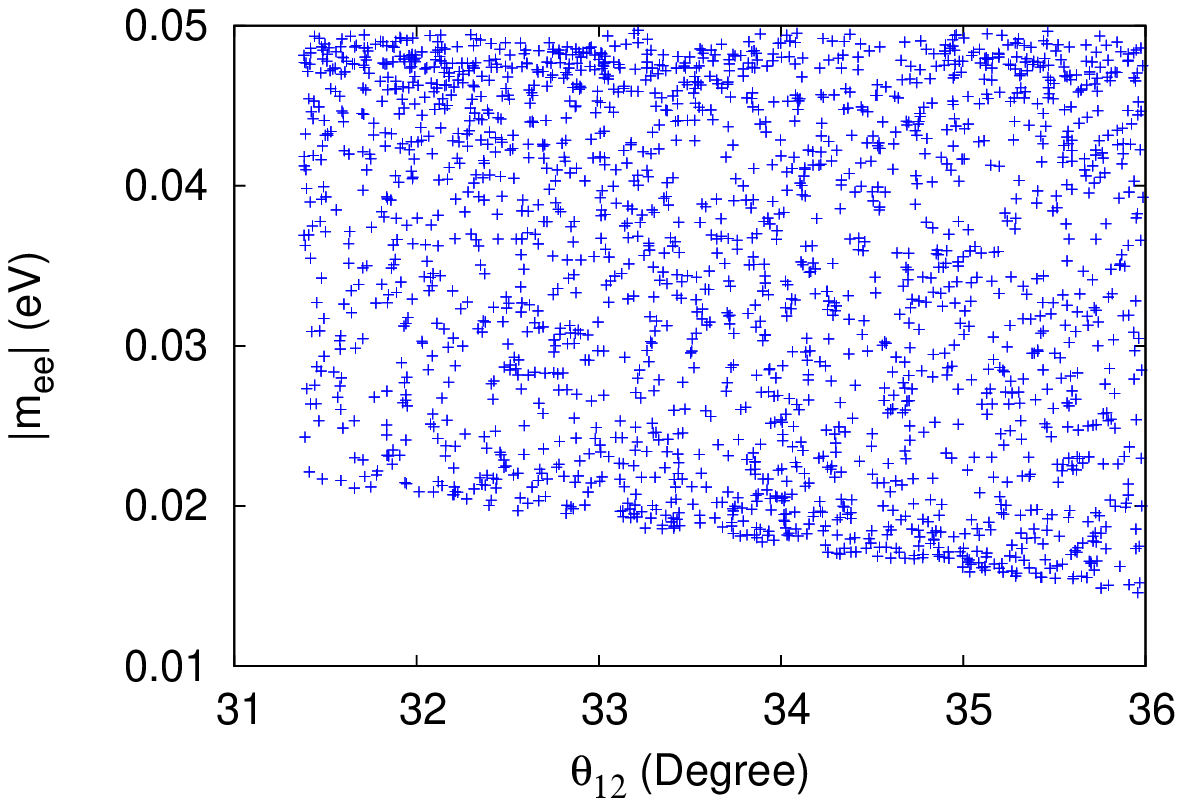, width=5.0cm, height=5.0cm}
\epsfig{file=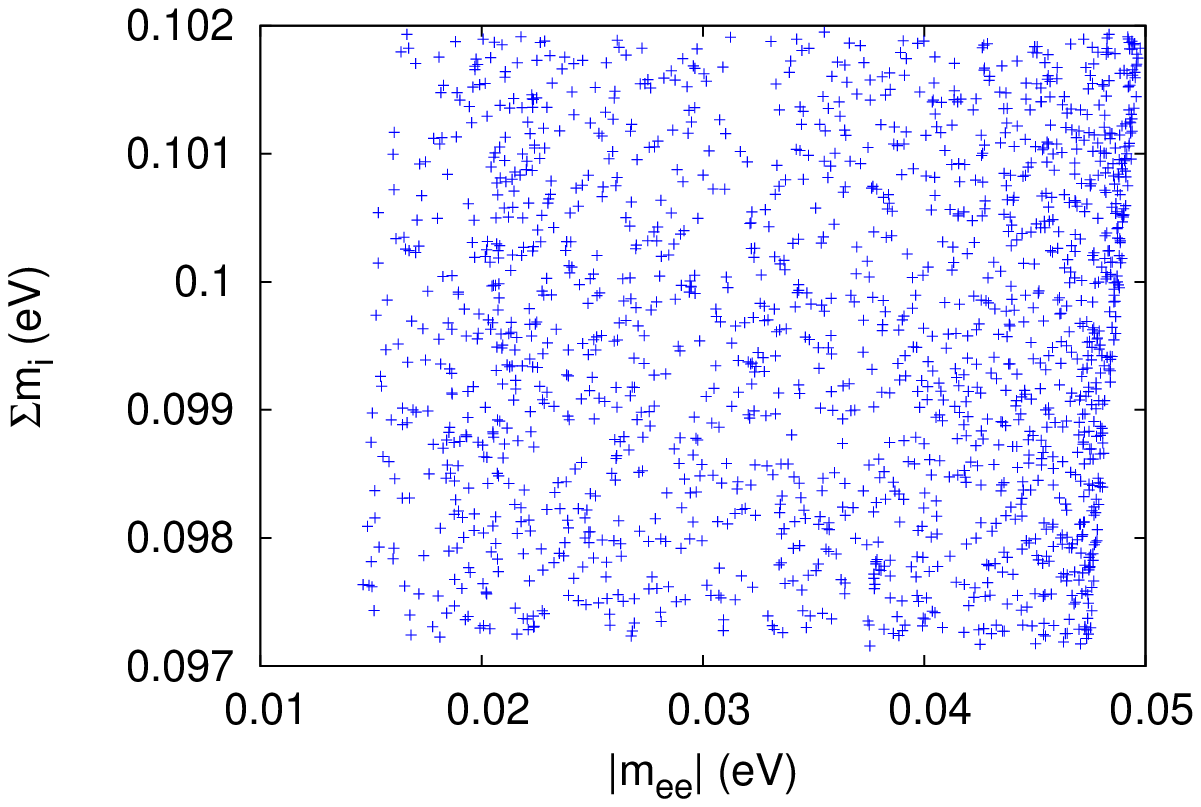, width=5.0cm, height=5.0cm}
\end{center}
\caption{Correlation plots between different parameters for NO (upper panel) and IO (lower panel) in texture IV-(D).}
\label{fig3}
\end{figure}

\begin{figure}[h]
\begin{center}
\epsfig{file=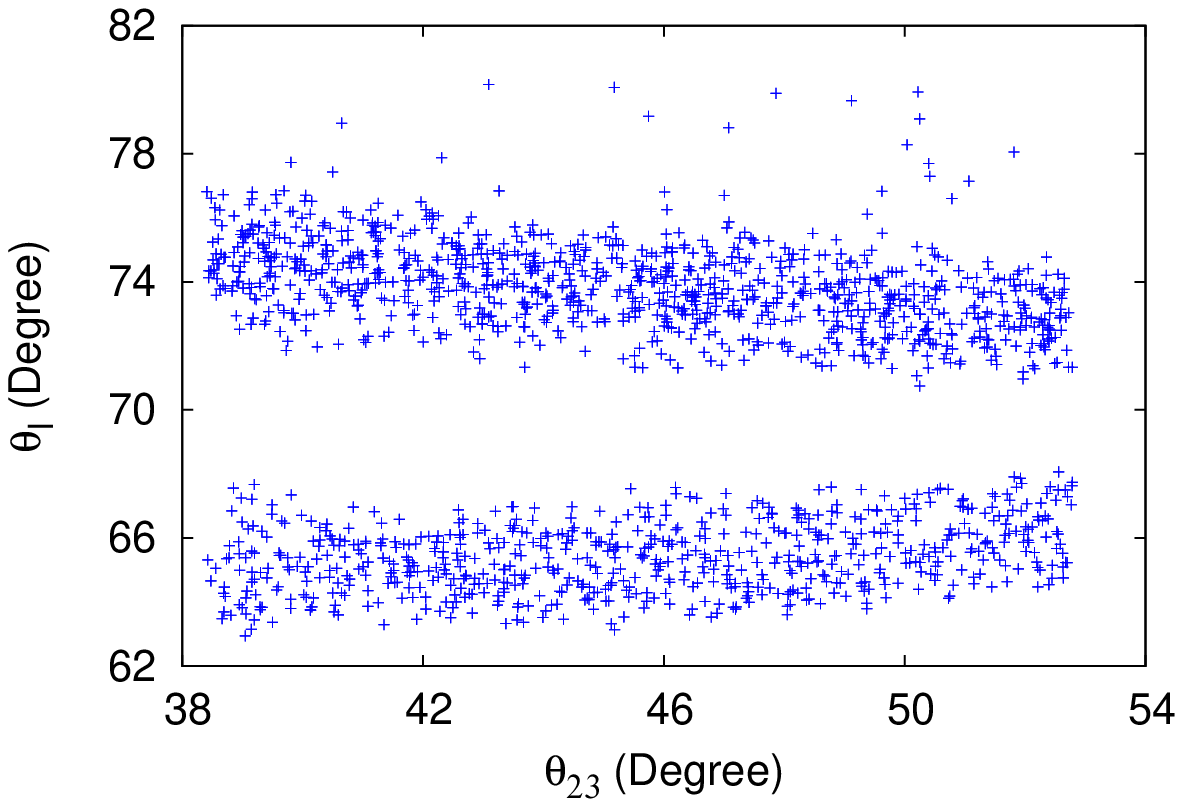, width=5.0cm, height=5.0cm}
\epsfig{file=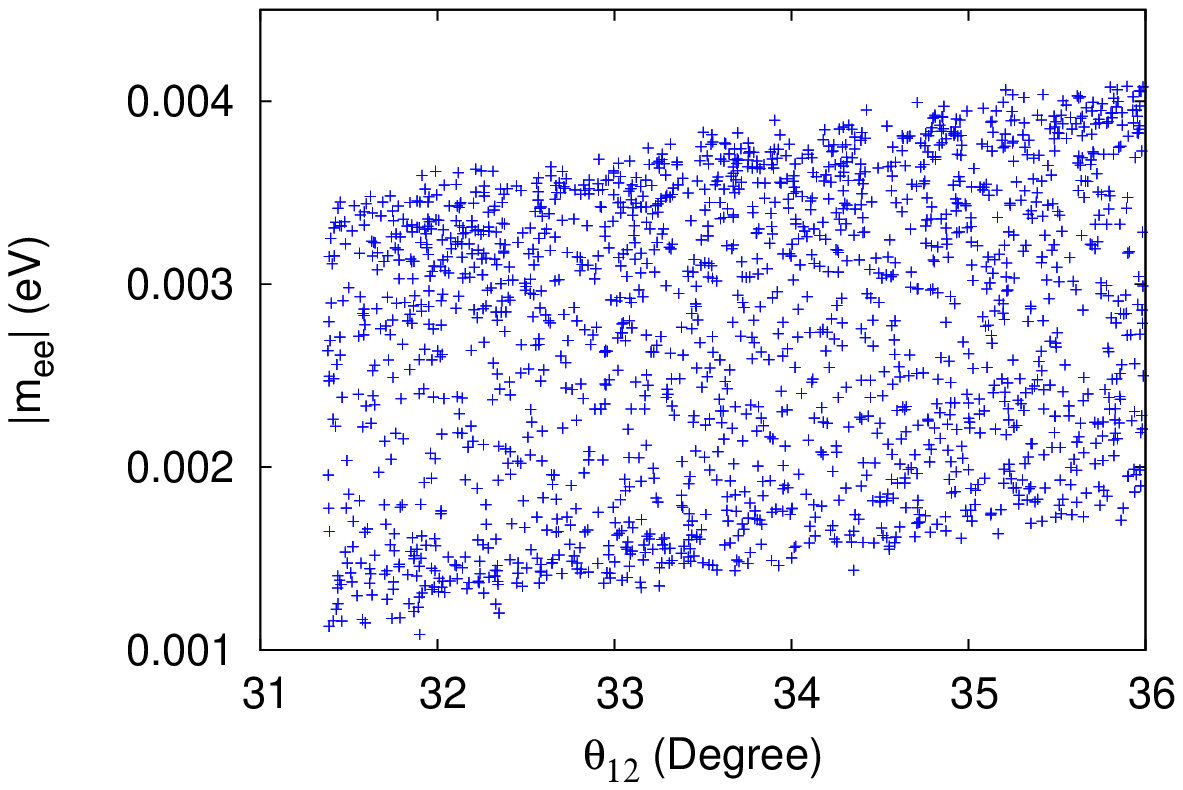, width=5.0cm, height=5.0cm}
\epsfig{file=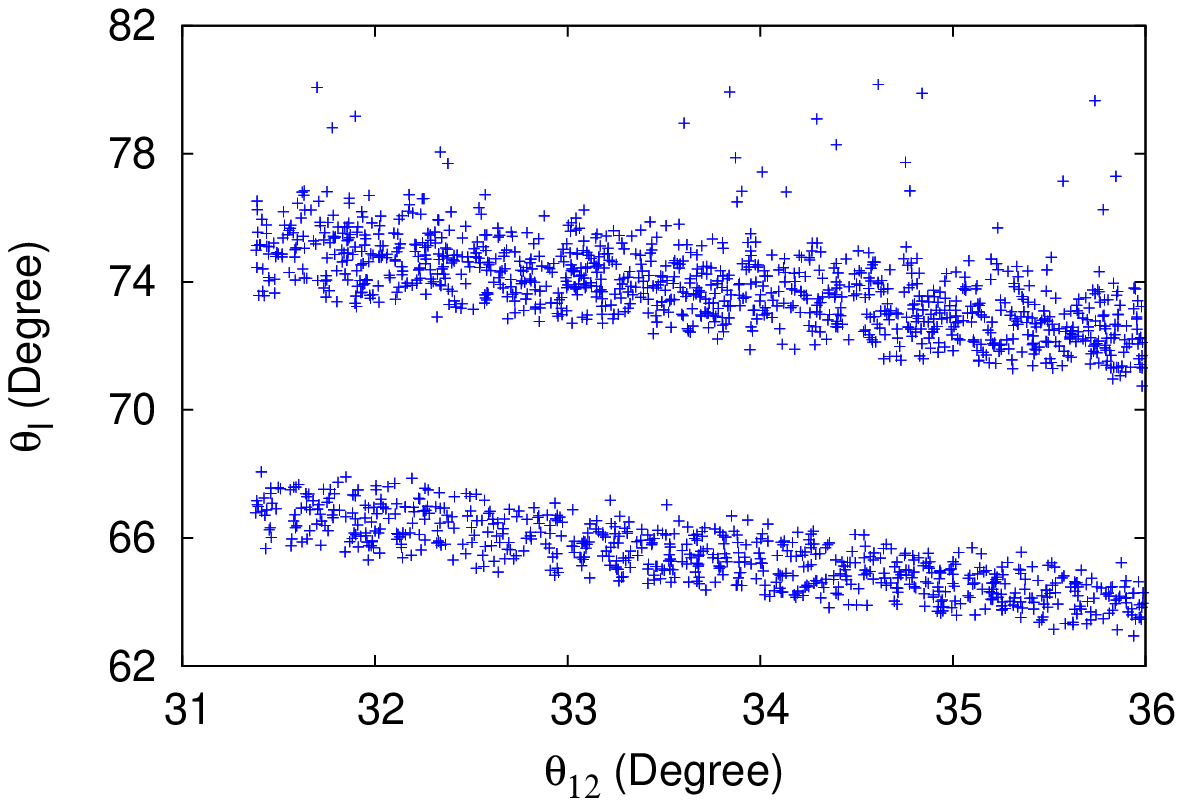, width=5.0cm, height=5.0cm}\\
\epsfig{file=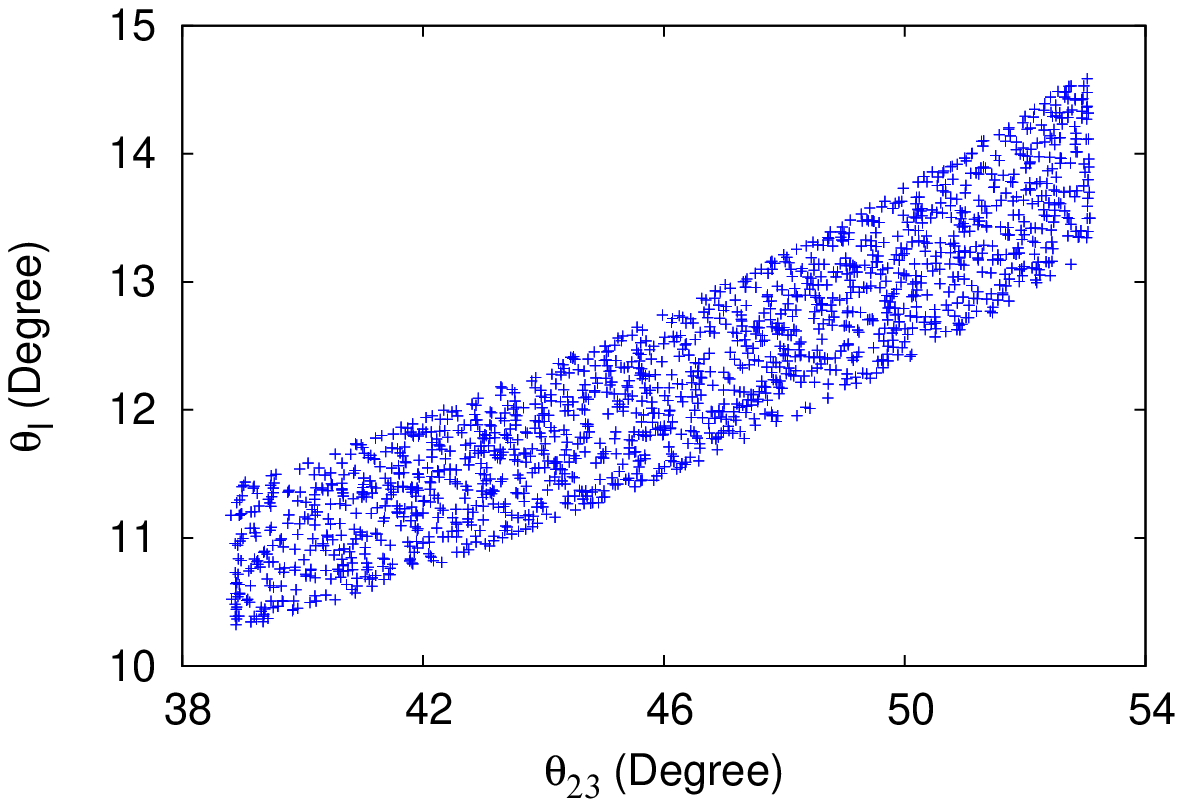, width=5.0cm, height=5.0cm}
\epsfig{file=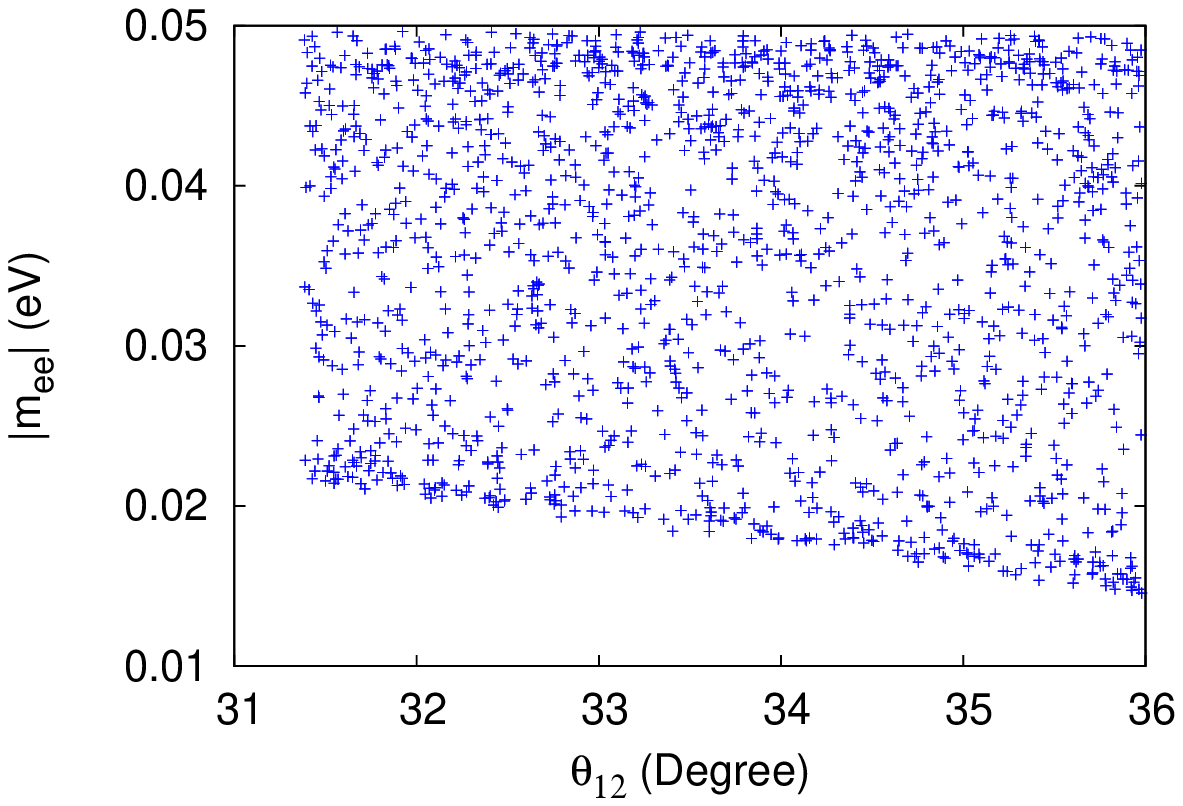, width=5.0cm, height=5.0cm}
\epsfig{file=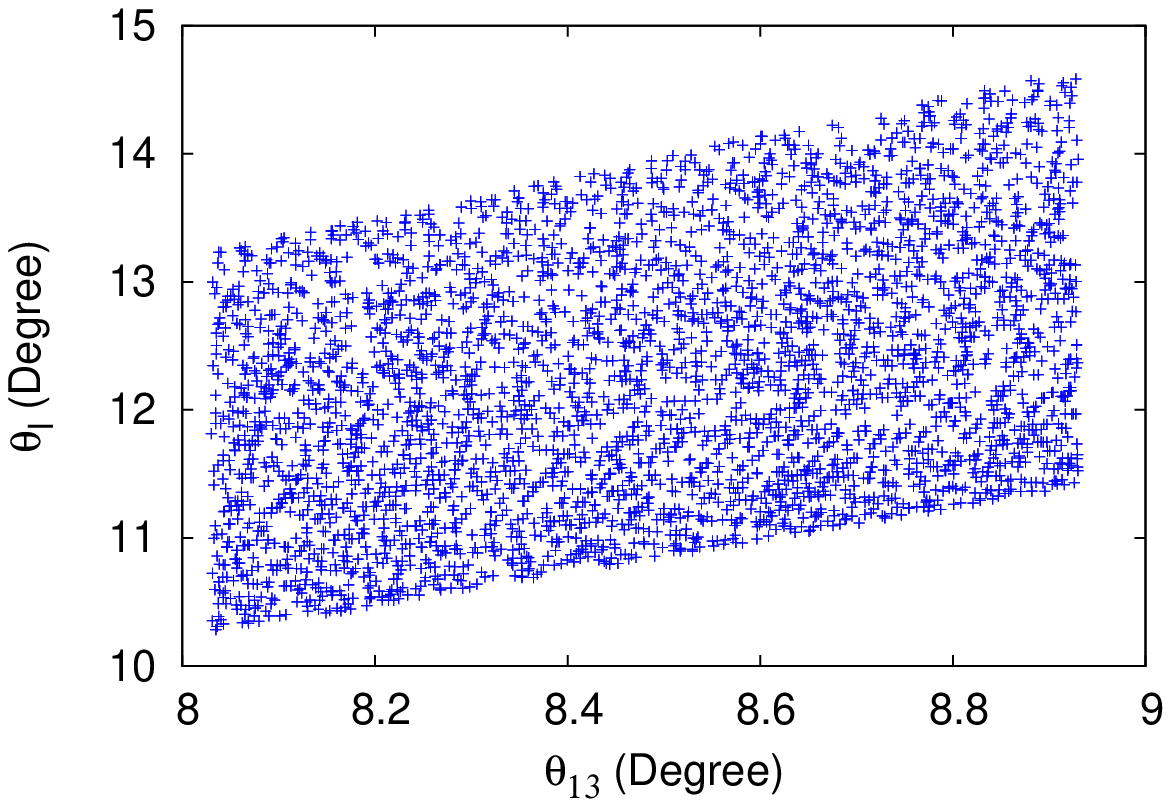, width=5.0cm, height=5.0cm}
\end{center}
\caption{Correlation plots between different parameters for NO (upper panel) and IO (lower panel) in texture IV-(E).}
\label{fig4}
\end{figure}

\begin{figure}[h]
\begin{center}
\epsfig{file=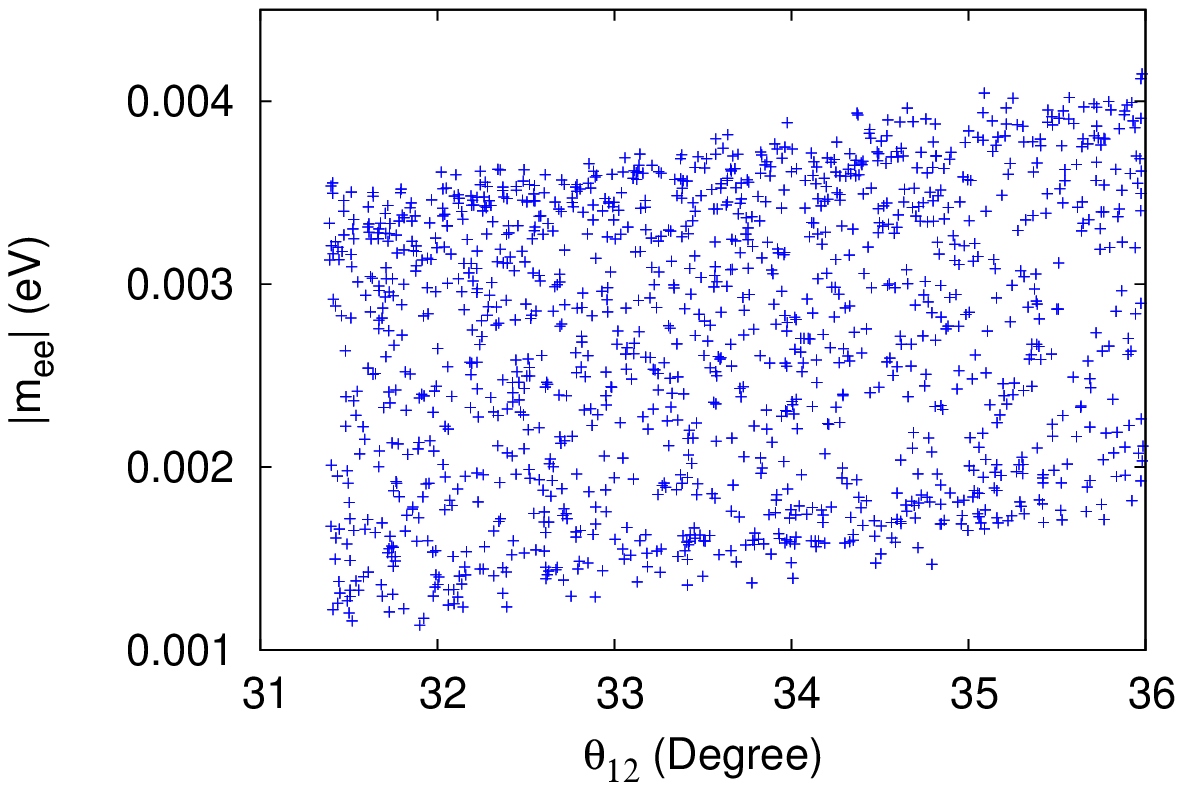,  width=5.0cm, height=5.0cm}
\epsfig{file=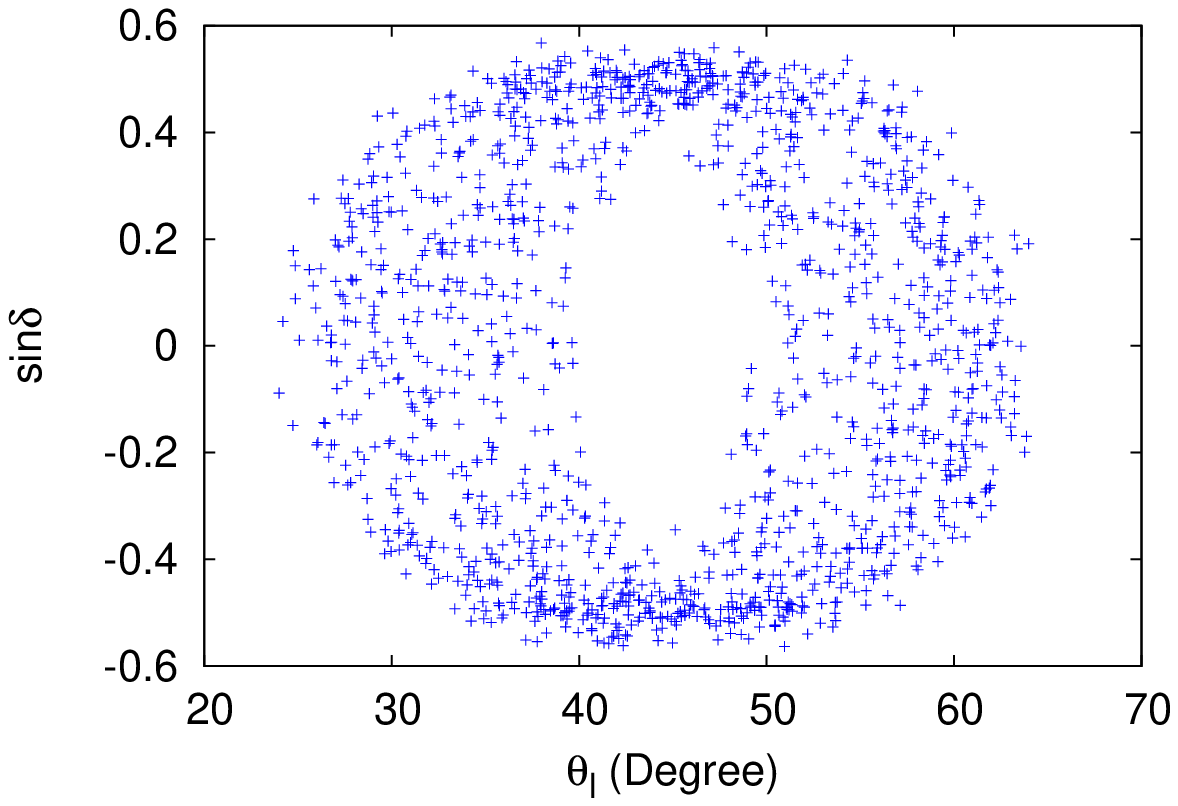,  width=5.0cm, height=5.0cm}
\epsfig{file=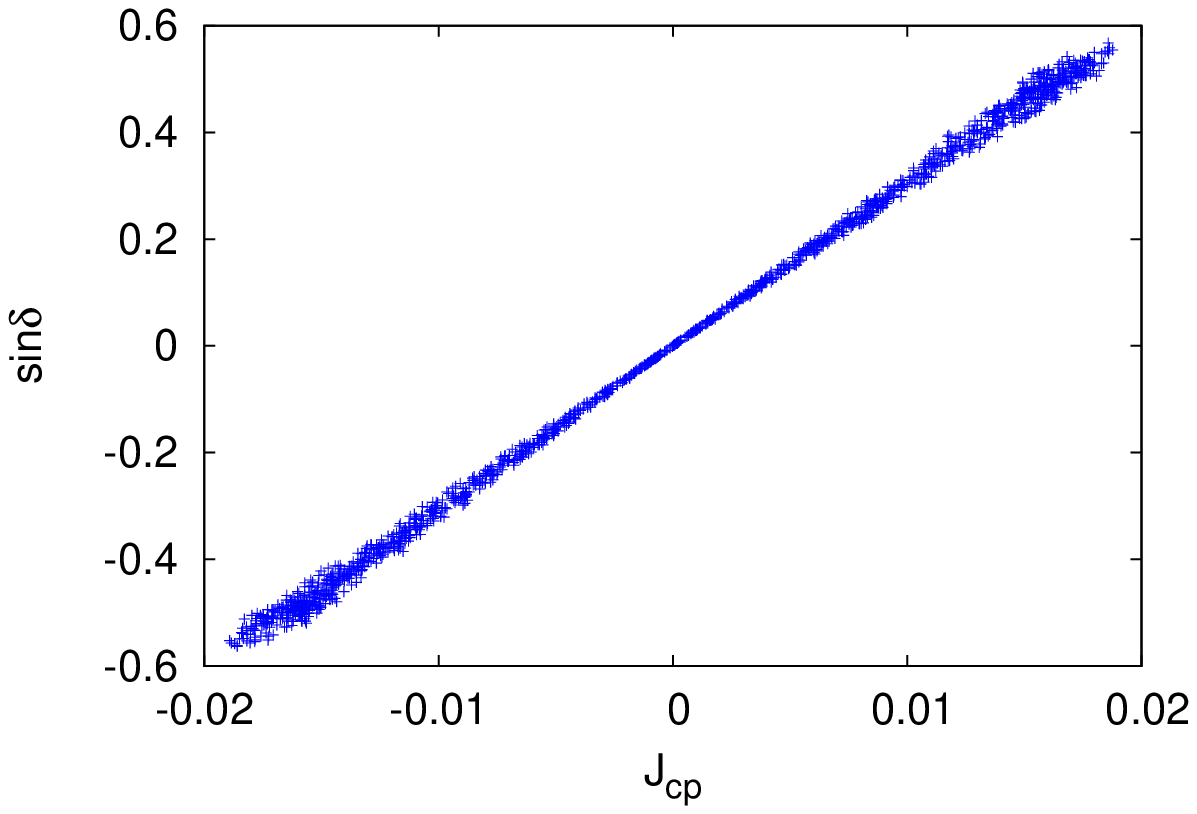,  width=5.0cm, height=5.0cm}\\
\epsfig{file=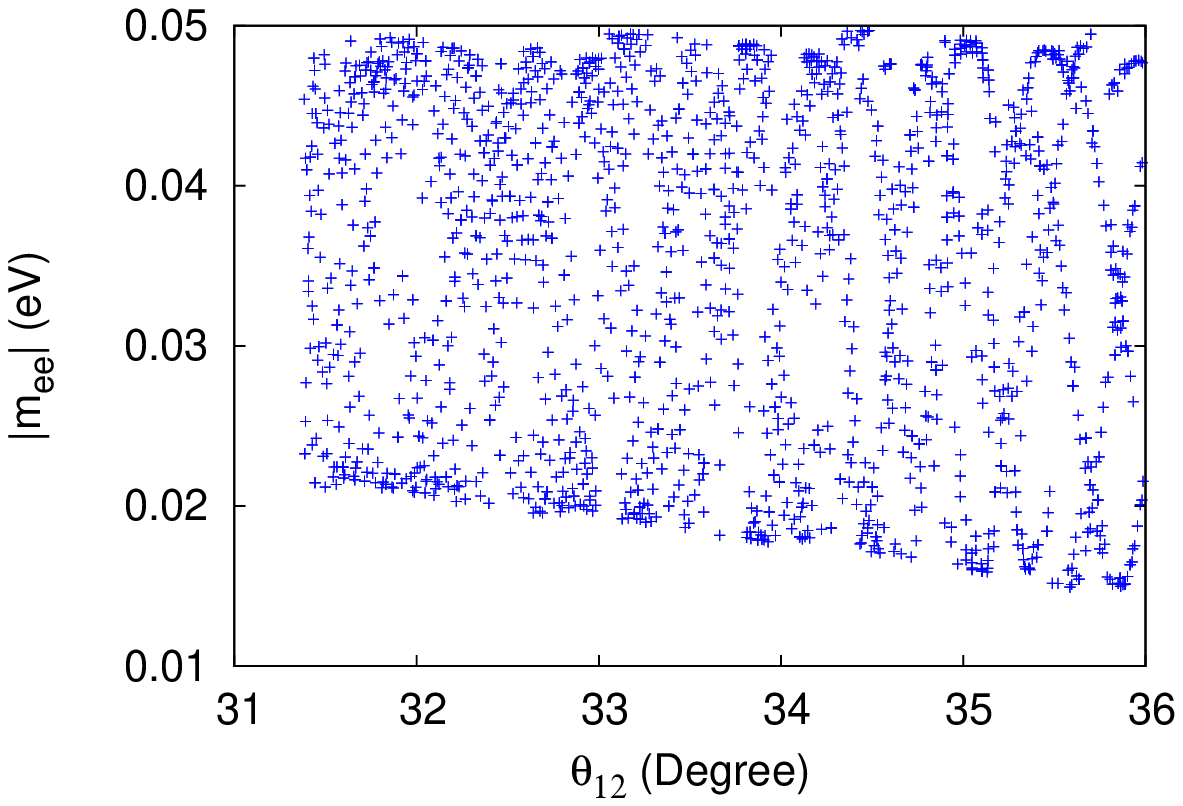,  width=5.0cm, height=5.0cm}
\epsfig{file=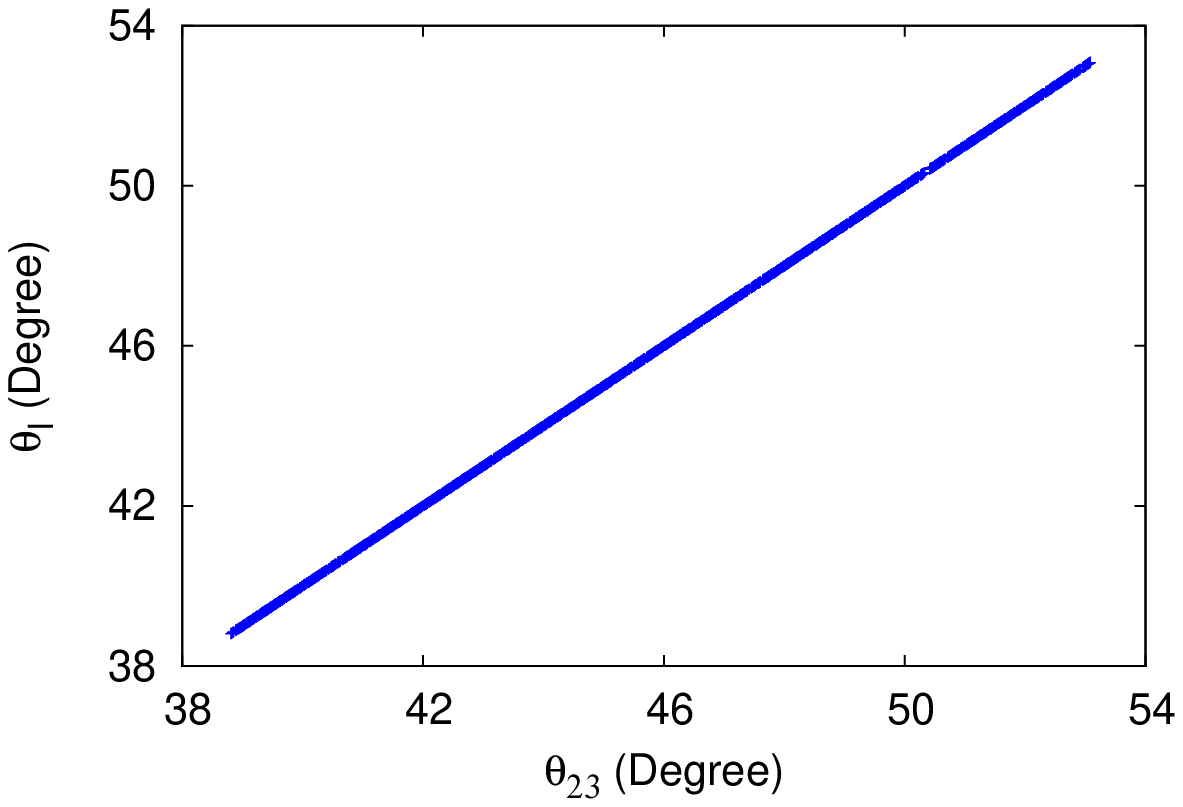,  width=5.0cm, height=5.0cm}
\epsfig{file=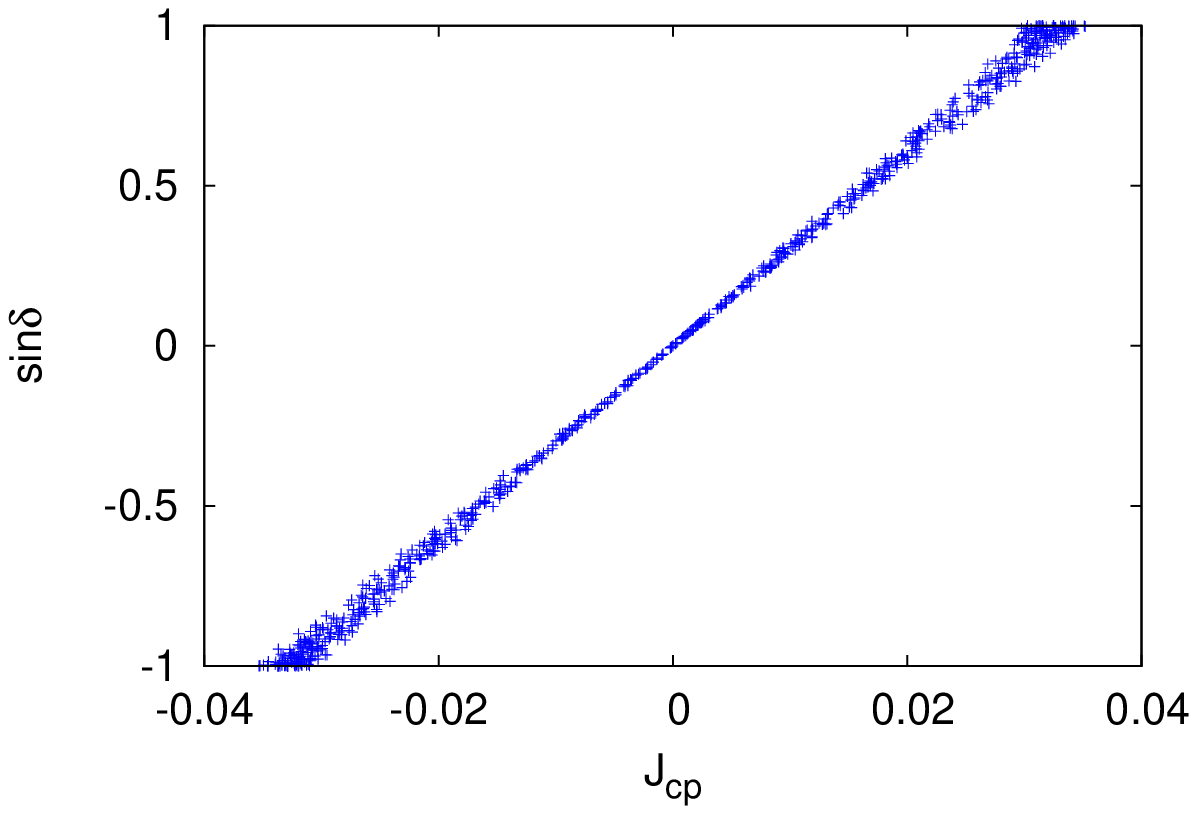,  width=5.0cm, height=5.0cm}
\end{center}
\caption{Correlation plots between different parameters for NO (upper panel) and IO (lower panel) in texture IV-(F).}
\label{fig5}
\end{figure}

\textbf{Texture IV-(F)}\\
In this texture structure, the neutrino mass matrix has three vanishing cofactors corresponding to (1,2), (2,2), and (2,3) positions and the charged lepton mass matrix has zero elements at (1,2) and (1,3) positions. $M_{\nu}$ and $H_{l}$ have the following structure:
\begin{eqnarray}
M_{\nu 16}=\left(
\begin{array}{ccc}
\times & \Delta & \times \\
\Delta & \Delta & \Delta \\
\times & \Delta & \times
\end{array}
\right)~~ \textrm{and}~~
H_{l1}=\left(
\begin{array}{ccc}
m_{e}^{2} & 0 & 0 \\
 0 & \times & \times \\
 0 & \times & \times
\end{array}
\right).
\end{eqnarray}\\
The neutrino mass matrix $M_{\nu 16}$ is related to $M_{\nu 14}$ by $S_{3}$ permutation symmetry as $M_{\nu 16}=S_{12} M_{\nu 14} S_{12}^{T}$, where $S_{12}$ is an element of the permutation group $S_{3}$. Therefore, the mixing matrix for structure $M_{\nu 16}$ is given by
\begin{eqnarray}
V_{\nu 16}=S_{12} V_{\nu 14}.
\end{eqnarray}
The PMNS mixing matrix for texture IV-(F) is given by
\begin{eqnarray}
U&=&V_{l1}^{\dagger} V_{\nu 16}=V_{l1}^{\dagger} S_{12} V_{\nu 14} \nonumber \\
&=& P^{\prime} \left(
\begin{array}{ccc}
0 & 1 & 0 \\ 
 \cos \theta_{l} & 0 & -e^{i \eta} \sin \theta_{l} \\
 \sin \theta_{l} & 0 & e^{i \eta} \cos \theta_{l}
\end{array}
\right) O_{\nu 14|NO(IO)}
\end{eqnarray}
where Eqs.(38) and (40) give $O_{\nu 14}$ for NO and IO, respectively. For IO of texture IV-(F), neutrino oscillation parameters are related as
\begin{equation}
\sin\theta_{13}=\frac{1}{\sqrt{c^{2}+1}},~~ \tan\theta_{23}= \tan\theta_{l}.
\end{equation}
The numerical results of Class-IV are presented in Figs. \ref{fig3}, \ref{fig4} and \ref{fig5} for both normal and inverted mass orderings. It can be seen from these figures that $\sin\delta$ spans the range ($-$1 - 1) except for NO of texture IV-(F) for which $\sin\delta$ is bounded by ($-$0.6 - 0.6). The Jarlskog CP invariant parameter $J_{CP}$ varies in the range ($-$0.02 - 0.02) for NO in texture IV-(F) whereas for other viable textures the range is ($-$0.036 - 0.036). There is a strong correlation between the charged lepton mixing angle $\theta_{l}$ and ($\theta_{12}, \theta_{23}$) for both mass orderings for all allowed textures of this class. Similar correlations are, also, present for $|m_{ee}|$ with $\theta_{12}$ and $\theta_{23}$ as shown in Figs. \ref{fig3}, \ref{fig4} and \ref{fig5}. For textures IV-(D), IV-(E) and IV-(F), the effective Majorana mass $|m_{ee}|$ is highly constrained to lie in the ranges (0.001 - 0.0045)eV for NO and (0.014 - 0.05)eV for IO. For viable textures of Class-IV, charged lepton correction $\theta_{l}$ is very large for normal mass ordering. For inverted mass ordering $\theta_{l}$ is small and the results are in agreement with Ref.\cite{dev2}, where charged lepton corrections were taken to be of the order of the Cabibbo angle. The sum of neutrino masses $\sum m_{i}$ varies in the range (0.057 - 0.061) eV for NO and (0.097 - 0.102) eV for IO. Textures IV-(A), IV-(B) and IV-(C) of Class-IV are phenomenologically incompatible with the $3\sigma$ neutrino oscillation data. 

\subsection*{Class-V}

In this class, there are three possible texture structures viz. $M_{\nu 17} H_{l1}, M_{\nu 17} H_{l2}$ and $M_{\nu 17} H_{l3}$. $H_{l}$ and $M_{\nu}$ for this class have the following form: 
\begin{eqnarray}
H_{l1}&=&\left(
\begin{array}{ccc}
m_{e}^2 & 0 & 0 \\
 0 & \times & \times \\
 0 & \times & \times
\end{array}
\right),
H_{l2}=\left(
\begin{array}{ccc}
\times & 0 & \times \\
 0 & m_{\mu}^2 & 0 \\
 \times & 0 & \times
\end{array}
\right),
H_{l3}=\left(
\begin{array}{ccc}
\times & \times & 0 \\
 \times & \times & 0 \\
 0 & 0 & m_{\tau}^2
\end{array}
\right),\\
M_{\nu 17}&=&\left(
\begin{array}{ccc}
\Delta & \times & \times \\
\times & \Delta & \times \\
\times & \times & \Delta
\end{array}
\right) \equiv
P_{\nu}\left(
\begin{array}{ccc}
 a & \sqrt{a b} & -\sqrt{a d} \\
 \sqrt{a b} & b & \sqrt{b d} \\
 -\sqrt{a d} & \sqrt{b d} & d \\
\end{array}
\right) P_{\nu}^T = P_\nu M_{\nu 17}^{r} P_\nu^T
\end{eqnarray}

where $P_{\nu}$ is a diagonal phase matrix. The real symmetric matrix $M_{\nu 17}^{r}$ can be diagonalized by the orthogonal matrix $O_{\nu 17}$:
\begin{equation}
M_{\nu 17}^{r}=O_{\nu 17} M_{\nu 17}^{D} O_{\nu 17}^{T}.
\end{equation} 
$O_{\nu 17}$ is given by
\begin{eqnarray}
O_{\nu 17}=\left(
\begin{array}{ccc}
 \frac{\sqrt{a d} \left(m_1-2 b\right)}{\left(a+b-m_1\right) m_1
   \sqrt{x+1}} & \frac{\sqrt{a d} \left(m_2-2
   b\right)}{\left(a+b-m_2\right) m_2
   \sqrt{y+1}} & \frac{\sqrt{a d} \left(m_3-2
   b\right)}{\left(a+b-m_3\right) m_3
   \sqrt{z+1}} \\
 \frac{\sqrt{b d} \left(2 a-m_1\right)}{\left(a+b-m_1\right) m_1
 \sqrt{x+1}} & \frac{\sqrt{b d} \left(2 a-m_2\right)}{\left(a+b-m_2\right) 
  m_2 \sqrt{y+1}} & \frac{\sqrt{b d} \left(2 a-m_3\right)}{\left(a+b-m_3\right) m_3   \sqrt{z+1}} \\
 \frac{1}{\sqrt{x+1}} & \frac{1}{\sqrt{y+1}} &
   \frac{1}{\sqrt{z+1}}
\end{array}
\right)
\end{eqnarray}
where 
\begin{eqnarray}
x=\frac{b d \left(2 a-m_1\right)^2}{m_1^2
   \left(a+b-m_1\right)^2}+\frac{a d
   \left(m_1-2 b\right)^2}{m_1^2
   \left(a+b-m_1\right)^2},\nonumber  \\
y=\frac{b d \left(2 a-m_2\right)^2}{m_2^2
   \left(a+b-m_2\right)^2}+\frac{a d^2
   \left(m_2-2 b\right)^2}{m_2^2
   \left(a+b-m_2\right)^2},  \\
z=\frac{b d \left(2 a-m_3\right)^2}{m_3^2
   \left(a+b-m_3\right)^2}+\frac{a d
   \left(m_3-2 b\right)^2}{m_3^2
   \left(a+b-m_3\right)^2} \nonumber
   \end{eqnarray}
and the parameters $a, b$ are related to neutrino masses $m_{1}, m_{2}$ and $m_{3}$ as
\begin{small}
\begin{eqnarray}     
a&=& \frac{-d^2+d (m_{1}+m_{2}+m_{3})+\sqrt{d \left(d
   (-d+m_{1}+m_{2}+m_{3})^2+m_{1} m_{2}
   m_{3}\right)}}{2 d}, \nonumber \\
b&=& -\frac{d^2-d (m_{1}+m_{2}+m_{3})+\sqrt{d \left(d
   (-d+m_{1}+m_{2}+m_{3})^2+m_{1} m_{2}
   m_{3}\right)}}{2 d}. \nonumber  
\end{eqnarray}
\end{small}
Three vanishing diagonal cofactors of $M_{\nu}$ relate mass eigenvalues as $m_{1} m_{2}+m_{2}m_{3}+m_{1}m_{3}=0$.
All texture structures of Class-V are inconsistent with the present neutrino oscillation data at $3\sigma$ level.\\
Texture structures III-(A) to III-(G) and V-(A) to V-(C) cannot simultaneously satisfy the experimental constraints on the mass squared differences and mixing angles and thus are inconsistent with neutrino oscillation data. In Table \ref{tab3} we have summarized the parameter space for the three mixing angles associated with each disallowed texture. One can see that the three mixing angles cannot simultaneously have values lying with in their experimental 3$\sigma$ ranges for the disallowed textures.
\begin{table}[h]
\begin{center}
\begin{tabular}{|c|c|c|c|c|c|c|}
 \hline
Texture & \multicolumn{3}{|c|}{NO} & \multicolumn{3}{|c|}{IO} \\
\cline{2-7}
  & $\theta_{13}$ & $\theta_{12}$ & $\theta_{23}$ & $\theta_{13}$ & $\theta_{12}$ & $\theta_{23}$\\ 
\hline
 III-(A)& $3 \sigma$ & $16^{\circ}$ - $20^{\circ}$ & $3 \sigma$ & $79^{\circ}$ - $90^{\circ}$ & $45^{\circ}$ - $90^{\circ}$ & $3 \sigma$ \\
III-(B)& $3 \sigma$ & $45^{\circ}$ - $46^{\circ}$ & $3 \sigma$ & $<7.5^{\circ}$ & $44^{\circ}$ - $46^{\circ}$ & $3 \sigma$  \\
 III-(C)& $3 \sigma$ & $45^{\circ}$ - $46^{\circ}$ & $3 \sigma$  & $<7.5^{\circ}$ & $44^{\circ}$ - $46^{\circ}$ & $3 \sigma$  \\
  III-(D)& $3 \sigma$ & $>50^{\circ}$ & $3 \sigma$ & $3 \sigma$ & $>50^{\circ}$ & $>83^{\circ}$  \\
 III-(E)& $3 \sigma$ & $3 \sigma$ & $16^{\circ}$ - $20^{\circ}$  & $3 \sigma$ & $44^{\circ}$ - $47^{\circ}$ & $<08^{\circ}$  \\
 III-(F)& $3 \sigma$ & $3 \sigma$ & $70^{\circ}$ - $74^{\circ}$  & $3 \sigma$ & $44^{\circ}$ - $47^{\circ}$ & $>82^{\circ}$  \\
  III-(G)& $3 \sigma$ & $>50^{\circ}$ & $3 \sigma$  & $3 \sigma$ & $>49^{\circ}$ & $<7^{\circ}$  \\
 \hline
 V-(A)& $53^{\circ}$ - $57^{\circ}$ & $3 \sigma$ & $3 \sigma$ & $35^{\circ}$ - $36^{\circ}$ & $3 \sigma$ & $3 \sigma$ \\
 V-(B)& $>20^{\circ}$ & $3 \sigma$ & $3 \sigma$ & $>20^{\circ}$ & $3 \sigma$ & $3 \sigma$ \\ 
 V-(C)& $>13^{\circ}$ & $3 \sigma$ & $3 \sigma$ & $>15^{\circ}$ & $3 \sigma$ & $3 \sigma$ \\
  \hline
 \end{tabular}
\caption{Parameter space for neutrino oscillation parameters for phenomenologically disallowed textures of classes III and V. 3$\sigma$ in the table denotes that the mixing angle lies within the 3$\sigma$ experimental range. }
\label{tab3}
\end{center}
\end{table}

\subsection*{Neutrino mass matrices with four vanishing cofactors}
Another possibility for lepton mass matrices includes four vanishing cofactors in the neutrino mass matrices with five non-zero elements in the charged lepton mass matrices. These textures have total eight degrees of freedom and such textures should have the same predictability as the two texture zero neutrino mass matrices in flavor basis. There are a total of 15 possible structures for $M_{\nu}$ having four vanishing cofactors. In a 3$\times$3 complex symmetric matrix, vanishing of any set of four cofactors leads to either the vanishing of the fifth or all six cofactors, simultaneously. A neutrino mass matrix where all six cofactors vanish leads to two degenerate neutrino masses which is incompatible with the experimental data. The remaining possible structures of $M_{\nu}$ which have five vanishing cofactors and non-degenerate mass eigenvalues are given below:
\begin{eqnarray}
\textrm{Class-VII}~~~~~~~~~M_{\nu 21}=\left(
\begin{array}{ccc}
\Delta & \Delta & \Delta \\
 \Delta & \Delta & \Delta \\
 \Delta & \Delta & \times
\end{array}
\right),
M_{\nu 22}=\left(
\begin{array}{ccc}
\Delta & \Delta & \Delta \\
 \Delta & \times & \Delta \\
 \Delta & \Delta & \Delta
\end{array}
\right),
M_{\nu 23}=\left(
\begin{array}{ccc}
\times & \Delta & \Delta \\
 \Delta & \Delta & \Delta \\
 \Delta & \Delta & \Delta
\end{array}
\right).
\end{eqnarray}
The charged lepton mass matrices with five non-zero matrix elements are of the following form:
\begin{eqnarray}
M_{l}=\left(
\begin{array}{ccc}
\times & 0 & 0 \\
 0 & \times & \times \\
 \times & 0 & \times
\end{array}
\right)
\end{eqnarray} 
with all possible reorderings of rows and columns of $M_{l}$. The Hermitian products $H_{l}=M_{l} M_{l}^{\dagger}$, whose diagonalization gives charged lepton mixing matrix $V_{l}$, are given below:
\begin{eqnarray}
H_{l1}=\left(
\begin{array}{ccc}
\times & 0 & 0 \\
 0 & \times & \times \\
 0 & \times & \times
\end{array}
\right),
H_{l2}=\left(
\begin{array}{ccc}
\times & 0 & \times \\
 0 & \times & 0 \\
 \times & 0 & \times
\end{array}
\right),
H_{l3}=\left(
\begin{array}{ccc}
\times & \times & 0 \\
 \times & \times & 0 \\
 0 & 0 & \times
\end{array}
\right),\\
H_{l4}=\left(
\begin{array}{ccc}
\times & \times & 0 \\
 \times & \times & \times \\
 0 & \times & \times
\end{array}
\right),
H_{l5}=\left(
\begin{array}{ccc}
\times & 0 & \times \\
 0 & \times & \times \\
 \times & \times & \times
\end{array}
\right),
H_{l6}=\left(
\begin{array}{ccc}
\times & \times & \times \\
 \times & \times & 0 \\
 \times & 0 & \times
\end{array}
\right),\\
H_{l7}=\left(
\begin{array}{ccc}
\times & \times & \Delta \\
 \times & \times & \times \\
 \Delta & \times & \times
\end{array}
\right),
H_{l8}=\left(
\begin{array}{ccc}
\times & \Delta & \times \\
 \Delta & \times & \times \\
 \times & \times & \times
\end{array}
\right),
H_{l9}=\left(
\begin{array}{ccc}
\times & \times & \times \\
 \times & \times & \Delta \\
 \times & \Delta & \times
\end{array}
\right)~.
\end{eqnarray}
where $\Delta$ at $ij$ position represents the vanishing cofactors corresponding to $ij$ elements.\\
If $M_{\nu}$ is one of Eq.(57) and $H_{l}$ is one of Eq.(59), the lepton mixing matrix $U$ has one of its elements equal to zero which is inconsistent with current experimental data and, hence, the charged lepton mass matrices listed in Eq.(59) are phenomenologically ruled out for four vanishing cofactors in $M_{\nu}$. Therefore, we focus on other forms of $H_{l}$ given in Eqs.(60) and (61). There are 18 possible combinations of $M_{\nu}$ and $H_{l}$ but not all are independent as interchanging rows and columns of $M_{\nu}$ is equivalent to reordering the rows and columns of $M_{l}$. Table \ref{tab4} gives independent combinations of $M_{\nu}$ and $H_{l}$ along with their viabilities. 
\begin{table}[h]
\begin{center}
\begin{tabular}{|c|c|c|c|}
 \hline
Class & Texture & $H_{l}$ $M_{\nu}$ & Viability \\
&   &  & NO IO \\
 \hline 
  & VII-(A)& $H_{l4}$, $M_{\nu21} \sim H_{l5}$, $M_{\nu 23} \sim H_{l6}$, $M_{\nu 22}$ & $\surd~~~\surd$ \\
 
 & VII-(B)& $H_{l4}$, $M_{\nu22} \sim H_{l5}$, $M_{\nu 21} \sim H_{l6}$, $M_{\nu 23}$ & $\surd~~~\surd$ \\
 
 & VII-(C)& $H_{l4}$, $M_{\nu23} \sim H_{l5}$, $M_{\nu 22} \sim H_{l6}$, $M_{\nu 21}$ & $\surd~~~\surd$  \\

 VII & VII-(D)& $H_{l7}$, $M_{\nu21} \sim H_{l8}$, $M_{\nu 23} \sim H_{l9}$, $M_{\nu 22}$ & $\times~~~\times$ \\
 
 & VII-(E)& $H_{l7}$, $M_{\nu22} \sim H_{l8}$, $M_{\nu 21} \sim H_{l9}$, $M_{\nu 23}$ & $\surd~~~\surd$ \\
 
 & VII-(F)& $H_{l7}$, $M_{\nu23} \sim H_{l8}$, $M_{\nu 22} \sim H_{l9}$, $M_{\nu 23}$ & $\surd~~~\surd$ \\
  \hline 
   \end{tabular}
\caption{Possible independent structures of $H_{l}, M_{\nu}$ and their viability.}
\label{tab4}
\end{center}
\end{table}
The neutrino mass matrices for this class cannot be diagonalized directly due to the presence of a non-removable phase and, hence, we diagonalize the Hermitian product $M_{\nu} M_{\nu}^{\dagger}$, which gives neutrino mixing matrix $V_{\nu}$. The neutrino mass matrix $M_{\nu}$ can be written as 
\begin{eqnarray}
M_{\nu 21}&=&\left(
\begin{array}{ccc}
\Delta & \Delta & \Delta \\
\Delta & \Delta & \Delta \\
\Delta & \Delta & \times
\end{array}
\right) \equiv
\left(
\begin{array}{ccc}
 A e^{i \phi_{1}} & B e^{i \phi_{3}}  & 0 \\
 B e^{i \phi_{3}} & D e^{i \phi_{2}} & 0 \\
 0 & 0 & 0
\end{array}
\right)
\end{eqnarray}
and the corresponding Hermitian matrix is given by
\begin{eqnarray}
M_{\nu 21} M^{\dagger}_{\nu 21}=\left(
\begin{array}{ccc}
 a & b e^{i \phi } & 0 \\
 b e^{-i \phi } & d & 0 \\
 0 & 0 & 0
\end{array}
\right)= P^{\dagger}_{\nu} \left(
\begin{array}{ccc}
 a & b & 0 \\
 b & d & 0 \\
 0 & 0 & 0
\end{array}
\right)
P_{\nu}
\end{eqnarray}
where 
\begin{eqnarray}
a &=& A^2+B^2,\\ \nonumber
b e^{i \phi } &=& A B e^{-i(\phi_{3}-\phi_{1})}+B D e^{i (\phi_{3}-\phi_{2})},\\
d &=& B^2+D^2 \nonumber
\end{eqnarray}
and $P_{\nu}=$ diag$(e^{i \phi },1,1)$.
For NO, the neutrino masses are $m_{1}=0$, $m_{2}=\sqrt{\Delta m^{2}_{21}}$, $m_{3}=\sqrt{\Delta m^{2}_{31}}$ and for IO $m_{1}=\sqrt{\Delta m^{2}_{23}-\Delta m^{2}_{21}}$, $m_{2}=\sqrt{\Delta m^{2}_{23}}$, $m_{3}=0$. The orthogonal matrix $O_{\nu21}$ for NO and IO is given by
\begin{eqnarray}
O_{\nu 21|NO}=
\left(
\begin{array}{ccc}
 0 & -\frac{\sqrt{d-m_2^2}}{\sqrt{m_3^2-m_2^2}} &
   \frac{\sqrt{m_3^2-d}}{\sqrt{m_3^2
   -m_2^2}} \\
 0 & \frac{\sqrt{m_3^2-d}}{\sqrt{m_3^2-m_2^2}} &
   \frac{\sqrt{d-m_2^2}}{\sqrt{m_3^2-m_2^2}} \\
 1 & 0 & 0 \\
\end{array}
\right)~~\textrm{and}~~
O_{\nu 21|IO}=
\left(\begin{array}{ccc}
 -\frac{\sqrt{d-m_1^4}}{\sqrt{m_2^4-m_1^4}} &
 \frac{\sqrt{m_2^4-d}}{\sqrt{m_2^4 -m_1^4}} & 0 \\
\frac{\sqrt{m_2^4-d}}{\sqrt{m_2^4 -m_1^4}} &
\frac{\sqrt{d-m_1^4}}{\sqrt{m_2^4-m_1^4}} & 0 \\
 0 & 0 & 1 \\
\end{array}
\right)
\end{eqnarray}
where $m_{2}^{2}<d<m_{3}^{2}$ for NO and $m_{1}^{2}<d<m_{2}^{2}$ for IO.\\
The mixing matrix for $M_{\nu 21}$ is given by
\begin{eqnarray}
V_{\nu 21}&=&P_{\nu} O_{\nu 21}.
\end{eqnarray}
For texture $M_{\nu22}$, which is related to $M_{\nu21}$ by permutation symmetry: $M_{\nu22}\rightarrow S_{23}M_{\nu21}S_{23}^{T}$, neutrino mixing matrix is given by  
\begin{eqnarray}
V_{\nu 22}&=&S_{23} P_{\nu} O_{\nu 21}.
\end{eqnarray}
For $M_{\nu23}$, neutrino mixing matrix is given by
\begin{eqnarray}
V_{\nu 23}&=&S_{123} P_{\nu} O_{\nu 21}.
\end{eqnarray}
The charged lepton mixing matrix for structures given in Eqs.(60) and (61) can be parametrized as
\begin{eqnarray}
V_{l}=P_{l} O_{l},~~\textrm{with}~~
O_{l}=\left(
\begin{array}{ccc}
 \overline{c}_{12} \overline{c}_{13} & \overline{c}_{13} \overline{s}_{12} & \overline{s}_{13} \\
 -\overline{c}_{23} \overline{s}_{12}- \overline{c}_{12} \overline{s}_{13} \overline{s}_{23} & \overline{c}_{12} \overline{c}_{23}- \overline{s}_{12} \overline{s}_{13} \overline{s}_{23} & \overline{c}_{13} \overline{s}_{23} \\
 \overline{s}_{12} \overline{s}_{23}- \overline{c}_{12} \overline{c}_{23} \overline{s}_{13} & - \overline{c}_{23} \overline{s}_{12} \overline{s}_{13}-\overline{c}_{12} \overline{s}_{23} & \overline{c}_{13} \overline{c}_{23}
\end{array}
\right)
\end{eqnarray}
where $\overline{s}_{ij}=\sin \chi_{ij}, \overline{c}_{ij}=\cos \chi_{ij}$ and $P_{l}=$ diag$(e^{i \phi^{'} },1,e^{i \phi^{''}})$ is unitary phase matrix.\\
For charged lepton mass matrix $H_{l4}$, a zero at the (1,3) position implies
\begin{eqnarray}
m_{e}^2 O_{l 11} O_{l31}+ m_{\mu}^2 O_{l12} O_{l32}+ m_{\tau}^2 O_{l13} O_{l33}=0
\end{eqnarray}
which gives
\begin{equation}
\sin\chi_{13}=\frac{\left(m_{e}^2-m_{\mu}^2\right) \sin \chi_{12} \cos\chi_{12} \tan \chi_{23})}{m_{e}^2 \cos^2 \chi_{12}+m_{2}^2 \sin^2\chi_{12}-m_{\tau}^2}.
\end{equation}
For charged lepton mass matrix $H_{l7}$, vanishing cofactor at (1,3) position implies
\begin{equation}
H_{l}|_{21} H_{l}|_{32}-H_{l}|_{22} H_{l}|_{31}=0 
\end{equation}
which gives
\begin{equation}
\sin\chi_{13}=\frac{m_\tau^2
 \left(m_e^2-m_\mu^2\right) \sin 2\chi_{12} \tan \chi_{23}}{m_\tau^2 \left(m_e^2-m_\mu^2\right) \cos 2\chi_{12}+m_e^2 \left(2 m_\mu^2-m_\tau^2\right)-m_\mu^2 m_\tau^2}.
\end{equation}
The PMNS mixing matrix for this class is given by
\begin{eqnarray}
U&=&V_{l}^{\dagger} V_{\nu},\nonumber\\
&=&O_{l}^{T} P_{l}^{\dagger} P_{\nu} O_{\nu |NO(IO)},\nonumber \\
&=&O_{l}^{T} P O_{\nu |NO(IO)}, 
\end{eqnarray}
where $P=$ diag$(e^{i (\phi-\phi^{'})},1,e^{-i\phi^{''}})$. The orthogonal matrices $O_{\nu}$ and $O_{l}$ are given in Eqs.(65) and (69).\\ 
In our numerical analysis, the parameters $\chi_{12}, \chi_{23}, d, \phi, \phi^{'}$ and $\phi^{''}$ have been generated randomly and 3$\sigma$ experimental constraints on oscillation parameters have been used. We found that all textures of this class are viable except VII-(D) which is unable to fit neutrino oscillation data for both mass orderings. For texture VII-(D), $\theta_{12}$ becomes too large in case of NO and $\theta_{13}$ becomes very small for IO. The predictions of all viable textures VII-(A), VII-(B), VII-(C), VII-(E) and VII-(F) for oscillation parameters are very similar for NO and IO. The Jarlskog CP invariant parameter $J_{cp}$ for this class varies in the range ($-$0.04 - 0.04) and $\sin\delta$ spans the range ($-$1 - 1) for both mass orderings. The ranges for $|m_{ee}|$ are (0.0032 - 0.0042)eV and (0.01-0.05)eV for NO and IO, respectively. The parameter $\sum m_{i}$ lies within the ranges (0.057 - 0.0605)eV and (0.097 - 0.102)eV, for NO and IO, respectively.  

\section*{Symmetry realization}
It has been shown in Ref. \cite{grimus} that vanishing cofactors and texture zeros in the neutrino mass matrix can be realized with extended scalar sector by means of discrete Abelian symmetries. We present a type-I seesaw realization of three vanishing cofactors using discrete Abelian flavor symmetries in the non-flavor basis.\\                                                                                                                                                                  
Abelian group $Z_{6}$ can be used for symmetry realization of mass matrix textures of Class-III. To obtain texture structure III-(H), one of the simplest possibilities is to have the following structures for $M_{D}$, $M_{R}$ and $M_{l}$:
\begin{eqnarray}
M_{D}=\left(
\begin{array}{ccc}
\times & 0 & 0 \\
 0 & \times & 0 \\
 0 & 0 & \times
\end{array}
\right),~~
M_{R}=\left(
\begin{array}{ccc}
\times & \times & 0 \\
 \times & 0 & \times \\
 0 & \times & 0
\end{array}
\right),~~ \textrm{and}~~
M_{l}=\left(
\begin{array}{ccc}
\times & \times & 0 \\
 0 & \times & 0 \\
 0 & 0 & \times
\end{array}
\right).
\end{eqnarray}
Here, $M_{D}$ is diagonal and texture zeros in $M_{R}$ propagate as vanishing cofactors in the effective neutrino mass matrix $M_{\nu}$. In addition to the SM left-handed $SU(2)_L$ lepton doublets $D_{l L}= \left(
\begin{array}{c}
\nu_{lL} \\
 l_{L}
 \end{array}\right), \ (l=e,\mu,\tau)$ and the right-handed charged lepton $SU(2)_L$ singlets $l_{R}$, we introduce three right handed neutrinos $\nu_{l R}$. In the scalar sector, we need two $SU(2)_{L}$ Higgs doublets $\phi$'s and two scalar singlets $\chi$'s.\\
We consider the following transformation properties of various fields under $Z_{6}$ for texture III-(H):
\begin{eqnarray}
\overline{D}_{eL} &\rightarrow & \omega^5 \overline{D}_{eL}, ~~e_{R} \rightarrow \omega^{2} e_{R}, ~ \nu_{eR} \rightarrow \omega \nu_{eR}\\ \nonumber
\overline{D}_{\mu L}& \rightarrow & ~~ \overline{D}_{\mu L}, ~~~\mu_{R} \rightarrow \omega \mu_{R},~  \nu_{\mu R}\rightarrow \omega^5 \nu_{\mu R}\\
\overline{D}_{\tau L} & \rightarrow & \omega^{3} \overline{D}_{\tau L},~~ \tau_{R} \rightarrow \omega^{3} \tau_{R},  ~\nu_{\tau R} \rightarrow \omega^{2}\nu_{\tau R} \nonumber
\end{eqnarray}
with $\omega=e^{\frac{2 \pi i}{6}}$ as generator of $Z_{6}$ group. The bilinears $\overline{D}_{l L} l_{R}, \overline{D}_{l L} \nu_{l R}$ and $\nu_{l R} \nu_{l R}$ relevant for $M_{l}$, $M_{D}$ and $M_{R}$, respectively, transform as \\
\begin{eqnarray}
\overline{D}_{l L} l_{R} \sim \left(
\begin{array}{ccc}
\omega & 1 & \omega^{2} \\
 \omega^{2} & \omega & \omega^{3} \\
 \omega^{5} & \omega^{4} & 1
\end{array}
\right), 
~~\overline{D}_{l L} \nu_{l R} \sim \left(
\begin{array}{ccc}
1 & \omega^{4} & \omega \\
 \omega & \omega^5 & \omega^{2} \\
 \omega^4 & \omega^{2} & \omega^{5}
\end{array}
\right)
 ~~\textrm{and} ~~
 \nu_{l R} \nu_{l R} \sim \left(
\begin{array}{ccc}
\omega^{2} & 1 & \omega^{3} \\
 1 & \omega^{4} & \omega \\
 \omega^{3} & \omega & \omega^{4}
\end{array}
\right).
\end{eqnarray}
For $M_{R}$, (1,2) element is invariant under $Z_{6}$ and hence the corresponding mass term is directly present in the Lagrangian without any scalar field. However, (1,1) and (2,3) matrix elements require the presence of two scalar singlets $\chi_{1}$ and $\chi_{2}$ which transform under $Z_{6}$ as $\chi_{1} \rightarrow \omega^{4} \chi_{1}$ and $\chi_{2} \rightarrow \omega^{5} \chi_{2}$, respectively. The other entries of $M_{R}$ remain zero in the absence of any further scalar singlets. To achieve non diagonal charged lepton mass matrix $M_{l}$, two scalar Higgs doublets are needed which transform under $Z_{6}$ as $\phi_{1}\rightarrow \phi_{1}$, $\phi_{2}\rightarrow \omega^{5} \phi_{2}$. A diagonal $M_{D}$ is obtained since scalar Higgs doublets $\tilde{\phi_{j}}(\equiv \imath\sigma_{2}\phi_{j}^{\ast})$ transform under $Z_{6}$ as $\tilde{\phi_{1}}\rightarrow \tilde{\phi_{1}}$ and $\tilde{\phi_{2}}\rightarrow \omega \tilde{\phi_{2}}$.
Thus, the $Z_6$ invariant Yukawa Lagrangian for texture III-(H) is given by
\begin{align}
\mathcal{L}_{Y}& = -Y_{ee}^l \overline{D}_{eL} \phi_2 e_{R} - Y_{e \mu}^l \overline{D}_{eL} \phi_1 \mu_{R} - Y_{\mu \mu}^l \overline{D}_{\mu L} \phi_2 \mu_{R}- Y_{\tau \tau}^l \overline{D}_{\tau L} \phi_1 \tau_{R} - Y_{ee}^{D} \overline{D}_{eL} \tilde{\phi_1} \nu_{eR}  \nonumber \\ & \quad - Y_{\mu \mu}^{D} \overline{D}_{\mu L} \tilde{\phi_2} \nu_{\mu R} -Y_{\tau \tau}^{D} \overline{D}_{\tau L} \tilde{\phi_2} \nu_{\tau R} + \frac{Y_{ee}^{R}}{2} \nu_{eR}^T C^{-1} \nu_{eR} \chi_1 + \frac{M_{e \mu}^{R}}{2} (\nu_{e R}^T C^{-1} \nu_{\mu R} +\nu_{\mu R}^T C^{-1} \nu_{e R}) \nonumber \\ & \quad + \frac{Y_{\mu \tau}^{R}}{2} (\nu_{\mu R}^T C^{-1} \nu_{\tau R} + \nu_{\tau R}^T C^{-1} \nu_{\mu R}) \chi_2 + \ \textrm{H. c.}
\end{align}
For texture III-(I), the structures for $M_{D}$, $M_{R}, M_{l}$ are given by
\begin{eqnarray}
M_{D}=\left(
\begin{array}{ccc}
\times & 0 & 0 \\
 0 & \times & 0 \\
 0 & 0 & \times
\end{array}
\right),~~
M_{R}=\left(
\begin{array}{ccc}
\times & 0 &\times \\
 0 & 0 & \times \\
 \times & \times & 0
\end{array}
\right),~~
M_{l}=\left(
\begin{array}{ccc}
\times & 0 & \times \\
 0 & \times & 0 \\
 0 & 0 & \times
\end{array}
\right)
\end{eqnarray}
and the leptonic fields are required to transform as 
\begin{eqnarray}
\overline{D}_{eL} &\rightarrow & \omega^{5} \overline{D}_{eL}, ~e_{R} \rightarrow \omega^{2} e_{R}, ~\nu_{eR} \rightarrow \omega \nu_{eR}\\ \nonumber
\overline{D}_{\mu L}& \rightarrow & \omega^3 \overline{D}_{\mu L}, ~\mu_{R} \rightarrow \omega^3 \mu_{R},~ \nu_{\mu R}\rightarrow  \omega^2 \nu_{\mu R}\\
\overline{D}_{\tau L} & \rightarrow & \overline{D}_{\tau L},~~~~ \tau_{R} \rightarrow \omega \tau_{R},  ~~~\nu_{\tau R} \rightarrow \omega^{5}\nu_{\tau R} \nonumber
\end{eqnarray}
under $Z_{6}$. The bilinears $\overline{D}_{l L} l_{R}, \overline{D}_{l L} \nu_{l R}$ and $\nu_{l R} \nu_{l R}$ corresponding to  $M_{l}$, $M_{D}$ and $M_{R}$ transform as
\begin{eqnarray}
\overline{D}_{l L} l_{R} \sim \left(
\begin{array}{ccc}
\omega & \omega^{2} & 1 \\
\omega^{5} & 1 & \omega^{4} \\
 \omega^{2} & \omega^{3} & \omega
\end{array}
\right), 
~~\overline{D}_{l L} \nu_{l R} \sim \left(
\begin{array}{ccc}
1 & \omega & \omega^4 \\
 \omega^4 & \omega^5 & \omega^{2} \\
 \omega & \omega^{2} & \omega^{5}
\end{array}
\right)
 ~~\textrm{and} ~~
 \nu_{l R} \nu_{l R} \sim \left(
\begin{array}{ccc}
\omega^{2} & \omega^{3} & 1 \\
\omega^{3} & \omega^4 & \omega \\
1 & \omega & \omega^4 \\
\end{array}
\right)~.
\end{eqnarray}
   
The two scalar fields required for non-zero elements in $M_{R}$, transform as $\chi_{1} \rightarrow \omega^4 \chi_{1}$, $\chi_{2} \rightarrow \omega^{5} \chi_{2}$ under $Z_{6}$. The desired form of $M_{l}$ requires the Higgs fields to transform as $\phi_{1}\rightarrow \phi_{1}$ and $\phi_{2}\rightarrow \omega^5 \phi_{2}$. The scalar Higgs doublets acquire non-zero vacuum expectation value (VEV) at the electroweak scale, while scalar singlets acquire VEV at seesaw scale.\\
Similarly, the symmetry realization for Class-IV can also be achieved using $Z_{6}$ group. The transformation properties of leptonic and scalar fields under $Z_{6}$ group are given in Table \ref{tab5}.    

\begin{table}[h]
\begin{center}
\begin{tabular}{|c|c|c|c|c|c|c|}
\hline
Model & $M_{l},~~~~~~~~~~~~~~~~~~~~ M_{R},~~~~~~~~~~~~~~~~~~~~ M_{D}$ & $\overline{D}_{e L}, \overline{D}_{\mu L}, \overline{D}_{\tau L}$ & $e_{R},\mu_{R}, \tau_{R}$ & $\nu_{e R}, \nu_{\mu R}, \nu_{\tau R}$ & $\phi_{1}, \phi_{2}, \phi_{3}$ & $\chi $ \\
 \hline 
 
IV-(D) &  \begin{tiny}
$ \left(
\begin{array}{ccc}
\times & \times & 0 \\
 0 & \times & 0 \\
 0 & 0 & \times
\end{array}
\right), \left(
\begin{array}{ccc}
\times & 0 & 0 \\
 0 & \times & 0 \\
 0 & 0 & \times
\end{array}
\right), \left(
\begin{array}{ccc}
\times & 0 & \times \\
 0 & 0 & \times \\
 0 & 0 & \times
\end{array} \right)$ \end{tiny} & $1,~~ \omega,~~ \omega^{2}$ & $\omega^{2}, \omega, \omega^{4}$ & $1,~~\omega,~~ \omega^{4}$ & $1,\omega^{4}, \omega^{5}$& $\omega^{4}$\\
  
 \hline
 
 IV-(E) & \begin{tiny}
 $\left(
\begin{array}{ccc}
\times & 0 & \times \\
0 & \times & 0 \\
0 & 0 & \times
\end{array}
\right),\left(
\begin{array}{ccc}
\times & 0 & 0 \\
 0 & \times & 0 \\
 0 & 0 & \times
\end{array}
\right),\left(
\begin{array}{ccc}
\times & 0 & \times \\
 0 & 0 & \times \\
 0 & 0 & \times
\end{array} \right)$ \end{tiny} & $1,~~ \omega^{2},~~ \omega$ & $\omega^{2}, \omega^{4}, \omega$ & $1,~~\omega,~~ \omega^{4}$ & $1,\omega^{4}, \omega^{5}$& $\omega^{4}$\\
  
 \hline
 
IV-(F) & \begin{tiny}
$\left(
\begin{array}{ccc}
\times & 0 & 0 \\
 0 & \times & \times \\
 0 & 0 & \times
\end{array}
\right),\left(
\begin{array}{ccc}
\times & 0 & 0 \\
 0 & \times & 0 \\
 0 & 0 & \times
\end{array}
\right),\left(
\begin{array}{ccc}
0 & 0 & \times \\
 \times & 0 & \times \\
 0 & 0 & \times
\end{array} \right)$ \end{tiny} & $\omega^2,~~1,~~\omega$ & $\omega^{4},\omega^{2},\omega$ & $1,~~\omega,~~\omega^{4}$ & $1,\omega^{4}, \omega^{5}$& $\omega^{4}$\\
\hline 
\end{tabular}
\caption{Transformation properties of lepton and scalar fields under $Z_{6}$ for Class-IV.}
\label{tab5}
\end{center}
\end{table}

For four vanishing cofactors in $M_{\nu}$ and five non-zero elements in $M_{l}$, the symmetry realization of textures can be achieved by cyclic group $Z_{9}$. Table \ref{tab6} depicts the transformation properties of leptonic and scalar fields under $Z_{9}$ group for Class-VII.    

\begin{table}[h]
\begin{center}
\begin{tabular}{|c|c|c|c|c|c|c|}
\hline
Model & $M_{l}$,~~~~~~~~~~~~~~~~~~~~ $M_{R}$,~~~~~~~~~~~~~~~~~~~~ $M_{D}$ & $\overline{D}_{e L}, \overline{D}_{\mu L}, \overline{D}_{\tau L}$ & $e_{R},\mu_{R}, \tau_{R}$ & $\nu_{e R}, \nu_{\mu R}, \nu_{\tau R}$ & $\phi_{1}, \phi_{2}, \phi_{3}$ & $\chi_{1},\chi_{2} $ \\
 \hline 
 
VII-(A) &  \begin{tiny}
$ \left(
\begin{array}{ccc}
\times & \times & 0 \\
 0 & \times & \times \\
 0 & 0 & \times
\end{array}
\right), \left(
\begin{array}{ccc}
\times & 0 & 0 \\
 0 & \times & 0 \\
 0 & 0 & \times
\end{array}
\right), \left(
\begin{array}{ccc}
\times & 0 & \times \\
 0 & 0 & \times \\
 0 & 0 & 0
\end{array} \right)$ \end{tiny} & $\omega^2,~~ \omega^4, ~~\omega^{6}$ & $\omega^{7}, \omega^5, \omega^{3}$ & $1,~~\omega^4,~~ \omega^{7}$ & $1,\omega^{2}, \omega^{3}$& $\omega, \omega^{4}$\\
  
 \hline
 
VII-(B) &  \begin{tiny}
$ \left(
\begin{array}{ccc}
\times & \times & 0 \\
 0 & \times & \times \\
 0 & 0 & \times
\end{array}
\right), \left(
\begin{array}{ccc}
\times & 0 & 0 \\
 0 & \times & 0 \\
 0 & 0 & \times
\end{array}
\right), \left(
\begin{array}{ccc}
0 & \times & \times \\
 0 & 0 & 0 \\
 0 & 0 & \times
\end{array} \right)$ \end{tiny} & $\omega^5, ~~\omega^6,~~ \omega^{4}$ & $\omega^{4}, \omega, \omega^{3}$ & $1,~~\omega^4,~~ \omega^{7}$ & $1,\omega^{2}, \omega^{3}$& $\omega, \omega^{4}$\\
  
 \hline
 
VII-(C) &  \begin{tiny}
$ \left(
\begin{array}{ccc}
\times & \times & 0 \\
 0 & \times & \times \\
 0 & 0 & \times
\end{array}
\right), \left(
\begin{array}{ccc}
\times & 0 & 0 \\
 0 & \times & 0 \\
 0 & 0 & \times
\end{array}
\right), \left(
\begin{array}{ccc}
0 & 0 & 0 \\
 0 & \times & \times \\
 0 & 0 & \times
\end{array} \right)$ \end{tiny} & $\omega^6,~~ \omega^5,~~ \omega^{4}$ & $1, \omega, \omega^{2}$ & $1,~~\omega^4,~~ \omega^{7}$ & $1,\omega^{2}, \omega^{3}$& $\omega, \omega^{4}$\\
  
 \hline
 VII-(E) &  \begin{tiny}
$ \left(
\begin{array}{ccc}
\times & \times & 0 \\
 0 & \times & 0 \\
 0 & \times & \times
\end{array}
\right), \left(
\begin{array}{ccc}
\times & \times & 0 \\
 0 & \times & 0 \\
 0 & \times & \times
\end{array}
\right), \left(
\begin{array}{ccc}
0 & 0 & \times \\
 0 & 0 & 0 \\
 \times & 0 & \times
\end{array} \right)$ \end{tiny} & $\omega^4,~~ \omega,~~ \omega^{2}$ & $\omega^{2}, \omega^5, \omega^{7}$ & $1,~~\omega^4,~~ \omega^{7}$ & $1,\omega^{2}, \omega^{3}$& $\omega, \omega^{4}$\\
  
 \hline
 VII-(F) &  \begin{tiny}
$ \left(
\begin{array}{ccc}
\times & \times & 0 \\
 0 & \times & 0 \\
 0 & \times & \times
\end{array}
\right), \left(
\begin{array}{ccc}
\times & 0 & 0 \\
 0 & \times & 0 \\
 0 & 0 & \times
\end{array}
\right), \left(
\begin{array}{ccc}
0 & 0 & 0 \\
 \times & 0 & \times \\
 0 & 0 & \times
\end{array} \right)$ \end{tiny} & $\omega,~~ \omega^2, ~~\omega^{4}$ & $\omega^{8}, ~~\omega^5,~~ \omega^{2}$ & $1,\omega^4, \omega^{7}$ & $1,\omega^{2}, \omega^{3}$& $\omega, \omega^{4}$\\
  
\hline 
\end{tabular}
\caption{Transformation properties of lepton and scalar fields under $Z_{9}$ for Class-VII.}
\label{tab6}
\end{center}
\end{table}
\section*{Conclusion}

We have investigated some new texture structures for lepton mass matrices. We have studied texture structures with three (four) vanishing cofactors in the neutrino mass matrix $M_{\nu}$ with four (five) non-zero elements in the charged lepton mass matrix $M_{l}$. There are 3 possible structures for $H_{l}$ and 20 possible structures of $M_{\nu}$ grouped into Classes-I, II, III, IV, V and VI, for three vanishing cofactors in $M_{\nu}$ and four non-zero elements in $M_{l}$. It is found that among six classes only Class-III and IV are phenomenologically viable. We also found that there are 5 viable textures having four vanishing cofactors in $M_{\nu}$ with five non-zero elements in $M_{l}$. By using the recent global neutrino oscillation data and data from cosmological experiments, a systematic phenomenological analysis has been done for each viable texture. We, also, presented the symmetry realization for the allowed texture structures using discrete Abelian symmetries in the framework of type-I seesaw mechanism.

\acknowledgements{ R. R. G. acknowledges the financial support provided by Department of Science and Technology, Government of India under the Grant No. SB/FTP/PS-128/2013. The research work of S. D. is supported by the Council for Scientific and Industrial Research, Government of India, New Delhi vide grant No. 03(1333)/15/EMR-II. S. D. gratefully acknowledges the kind hospitality provided by IUCAA, Pune.}

\end{document}